\documentclass[a4paper,fleqn,usenatbib]{mnras}

\usepackage{newtxtext}
\usepackage{mathptmx}
\usepackage[T1]{fontenc}
\usepackage{ae,aecompl}

\usepackage{graphicx}	
\usepackage{amsmath}	
\usepackage{amssymb}	
\usepackage{microtype}

\newcommand{\D}{\mathrm{d}}
\newcommand{\Ms}{{\ensuremath{\mathrm{M}_{\odot}}}}

\title[On Pop III Supernova Rates
]{A New Statistical Model for Population III Supernova Rates: Discriminating Between $\Lambda$CDM and WDM Cosmologies}

\author[M. Magg et al.]{Mattis Magg$^{1}$\thanks{E-mail: mattis.magg@stud.uni-heidelberg.de}, 
Tilman Hartwig$^{2, 3}$, 
Simon C. O. Glover$^{1}$, 
Ralf S. Klessen$^{1, 4}$, 
\newauthor Daniel J. Whalen$^{5}$\\
$^{1}$Universit\"at Heidelberg, Zentrum f\"ur Astronomie, Institut f\"ur Theoretische Astrophysik, Albert-Ueberle-Str. 2,
\\D-69120 Heidelberg, Germany\\
$^{2}$Sorbonne Universit\'es, UPMC Univ Paris 06, UMR 7095, Institut d'Astrophysique de Paris, F-75014, Paris, France\\
$^{3}$CNRS, UMR 7095, Institut d'Astrophysique de Paris, F-75014, Paris, France\\
$^{4}$Universit\"{a}t Heidelberg, Interdiszipli\"{a}res Zentrum f\"{u}r Wissenschaftliches Rechnen\\
$^{5}$Institute of Cosmology and Gravitation, University of Portsmouth, Dennis Sciama Building, Portsmouth PO1 3FX, UK
}

\date{Accepted XXX. Received YYY; in original form ZZZ}

\pubyear{2016}

\begin{document}
\label{firstpage}
\pagerange{\pageref{firstpage}--\pageref{lastpage}}
\maketitle

\begin{abstract}
With new observational facilities becoming available soon, discovering and characterising supernovae from the first stars will open up alternative observational windows to the end of the cosmic dark ages. Based on a semi-analytical merger tree model of early star formation we constrain Population III supernova rates. We find that our method reproduces the Population III supernova rates of large-scale cosmological simulations very well. Our computationally efficient model allows us to survey a large parameter space and to explore a wide range of different scenarios for Population III star formation. Our calculations show that observations of the first supernovae can be used to differentiate between cold and warm dark matter models and to constrain the corresponding particle mass of the latter. Our predictions can also be used to optimize survey strategies with the goal to maximize supernova detection rates.
\end{abstract}




\section{INTRODUCTION}
The first stars appeared after redshift $z \sim$ 30, ending the cosmic dark ages and beginning the process of cosmological reionisation \citep{BrommReview,GloverReview,wan04,ket04,abs06,awb07,wet08b,wet10}. They also enriched the early cosmos with the first heavy elements
\citep{mbh03,byh03,ss07,wet08a,bsmith09,ritt12, KarlssonReview,ss13} and 
may be the origin of supermassive black holes today \citep[e.g.,][]{milos09a,milos09b,awa09,th09,pm11,pm12,jlj12a,agarw12,wf12,vol12,pm13,jet13,latif13a,latif13c,jet14}.

In spite of their importance for early structure formation, not much is known for certain about the properties of Population III (or Pop III) stars \citep{fsg09,fg11,dw12,GloverReview}. They formed in primordial gas which, due to its lack of metals, cools less efficiently than the interstellar medium (ISM) today. As a result, the pristine gas fragmented and collapsed on larger mass scales, so it is generally thought that the Pop III initial mass function (IMF) is top-heavy \citep{bl04} compared to later generations of stars. Numerical simulations of Pop III star formation remain inconclusive because no simulations have yet been able to follow the growth and evolution of a Pop III star or stellar cluster from its birth to the end of its life, while having sufficient resolution to fully resolve gravitational fragmentation in the Pop III accretion disk, and while also including all of the key physical processes (e.g. magnetic fields, radiative feedback, etc.), which set the stage for fragmentation of the accretion disc and determine when the stars eventually stop growing \citep{bcl99,abn00,abn02,bcl02,nu01,on07,y08,turk09,stacy10,clark11,sm11,Greif11b,hos11,get12,stacy12,susa13,hir13,Hartwig15a}. Despite these shortcomings, it is currently believed that Pop~III stars form in binaries or small-number, multiple stellar systems \citep[e.g.][]{turk09, clark11, sb13, Stacy16}, and that they do so with masses ranging from the sub-solar regime potentially up to several hundreds of solar masses. Some models suggest a logarithmically flat distribution of masses \citep{Greif11b, sm11, get12, Dopcke13, GloverReview} in contrast to the IMF observed today, which exhibits a peak in the subsolar regime followed by a power-law decline towards larger masses \citep{KroupaIMF, ChabrierIMF}. Top-heavy Pop III IMFs are supported by detailed investigations of the chemical abundances of extremely metal-poor stars in the Galactic halo, which suggest that Pop III stars in the mass range of 15 -- 40\,\Ms\ were responsible for much of the early enrichment \citep[e.g.][]{bc05, fet05, Frebel08, jet09b, aoki14, Chen16}.

The detection of Pop III stars is one the major goals of astronomy in the coming decade. Low-mass Pop III stars with masses below $\sim 0.8\,\Ms$ should have survived until the present day, and could potentially be detected in current and future Galactic archaeological surveys \citep{Ishiyama16}. This approach can be used to constrain the low-mass end of the Pop III IMF \citep[e.g., ][]{Hartwig15b}. Finding high-mass Pop III stars is more difficult, because they have short lifetimes and are only found at high redshifts. Furthermore, even the most massive primordial stars are too faint \citep{Schaerer2002} to be directly visible even with next-generation telescopes such as the \emph{James Webb Space Telescope} \citep[\emph{JWST}:][]{jwst06}, \emph{Euclid} \citep{euclid}, the \emph{Wide-Field Infrared Survey Telescope} \citep[\emph{WFIRST};][]{wfirst}, or the European Extremely Large Telescope \citep[E-ELT:][]{EELT07, EELT14}. The recent detection of the gravitational wave signal of the binary black hole merger GW150914 \citep{LIGO} opens a new but limited window on the first stars as mergers between very massive Pop III remnants could be detected as contribution to the statistical background (\citealt{Inayoshi16}, but \citealt{Dvorkin16}) and directly within a few decades \citep{Hartwig2016b}.

Detections of Pop III SNe in the near infrared (NIR) in the coming decade could probe the properties of the first stars, because the mass of the progenitor can be inferred from its light curve. The final fates of the first stars depend primarily on their masses and rotation rates \citep{HegerWoosley2002}. Pop III stars from 8 -- 40 \Ms\ die as core-collapse (CC) SNe with energies similar to those of such events today. Stars from 40 -- 90 \Ms\ directly collapse to a black hole (BH) with no visible explosion, except in the rare case that they are very rapidly rotating, when they can explode as hypernovae (HNe) or gamma ray bursts \citep[GRBs; e.g.,][]{gou04,mesz10,mes13a}. In a few cases ejecta from a CC SN can also crash into a dense shell ejected by the star a few years prior to its death. The collision produces an event that is very bright in the UV, a Type IIn SN \citep[e.g.,][]{nsmith07b,moriya12,Whalen13c}.

Above 90 \Ms\ Pop III stars can encounter the pair instability (PI) at the end of helium burning, in which e$^-$e$^+$ production in the core of the star causes it to contract, triggering explosive oxygen and silicon burning \citep{rs67,brk67,bet84}. In non-rotating stars from 90 -- 140 \Ms\ the PI does not totally disrupt the star but causes a series of mass ejections that can later collide and, like Type IIn SNe, produce very luminous events in the UV \citep[pulsational PI SNe, or PPI SNe; e.g.,][]{wbh07,wet13d,chen14a}. At 140 -- 260 \Ms\ the PI produces an explosion that completely unbinds the star and leaves no compact remnant \citep[e.g.,][]{HegerWoosley2002, jw11,chen14c}. PI SNe can have a hundred times the energy of a CC SN and can potentially be detected at very early epochs. Rotating Pop III stars from 90 -- 140\,\Ms\ shed their hydrogen envelopes and explode as bare He cores that are less luminous than 140 -- 260\,\Ms\ explosions \citep{cw12,cwc13,CW15}.

Recent numerical simulations of PI SN light curves show that 140 -- 260 \Ms\ PI SNe and 110 \Ms\ PPI SNe will be visible to \emph{JWST} and the E-ELT up to $z >$ 30 and to \emph{Euclid} and \emph{WFIRST} at $z \sim$ 10 -- 20 \citep{kasen11,hum12,pan12a,wet12a,wet12b}. 90 -- 140  \Ms\ PI SNe of rotating Pop III stars and 25 -- 50 \Ms\ HNe will be visible to \emph{JWST} and E-ELT at $z \sim$ 5 -- 15 and to \emph{Euclid} and \emph{WFIRST} at $z \sim$ 4 -- 5 \citep{smidt13a,smidt14a}. Pop III Type IIn SNe will be visible at $z \sim$ 20 to \emph{JWST} and E-ELT and at $z \sim$ 5 -- 10 to \emph{Euclid} and \emph{WFIRST} \citep{Whalen13c}. Pop III CC SNe, perhaps the most numerous type of explosion at high redshift, will be visible at $z \sim$ 10 -- 15 to \emph{JWST} and E-ELT and at $z \lesssim$ 7 to \emph{Euclid} and \emph{WFIRST} \citep{wet12c}.

Detection strategies for Pop III SNe require estimates of SN rates as a function of redshift \citep[e.g.,][]{hum12,ds13,ds14}. Large scale multi-physics cosmological simulations such as the recent First Billion Years Project (FiBY) \citep[e.g.][]{FiBY1, Paardekooper2013,agarw14}, Renaissance \citep{Xu2014,ren15} and the Birth of a Galaxy \citep{wise12,wise12a} campaigns can produce realistic SN rates over a range of redshifts but can require months to perform and cannot easily explore the full cosmological parameter space. They also cannot easily address the impact of new observational constraints on these rates, such as the newly revised optical depth to Thompson scattering, $\tau_e =$ 0.054 \citep{Planck2016}. We instead calculate Pop III SN rates with a detailed semi-analytic merger tree model that incorporates both radiative and SN feedback. Our model is computationally much less expensive than cosmological simulations and enables us to probe the impact of a variety of parameters on SN rates, such as the Pop III IMF. SN rates from our models in turn can be used to better constrain some of these uncertain parameters.

Where not stated otherwise we assume a flat $\Lambda$CDM Universe. Except for $\tau_e$, which is treated separately, we use the best fit cosmological parameters from \citet{Planck2013} ($H_0= 67.77\,\mathrm{km}\,\mathrm{s}^{-1}\mathrm{Mpc}^{-1},\ \Omega_m=0.3086,\ \Omega_b=0.04825,\ \Omega_{\Lambda}=0.6914,\ \sigma_8=0.8288,\ n_s=0.9611$). In Section \ref{sect:method} we describe our merger tree model for computing SN rates. In Section \ref{sect:res} we calculate Pop III SN rates for a variety of Pop III IMFs and cosmological parameters including dark matter model. We discuss caveats to our results in Section \ref{sect:caveat} and we conclude in Section \ref{sect:conc}.

\section{NUMERICAL METHOD}
\label{sect:method}
To calculate cosmic SN rates, we have improved the semi-analytical model of early star formation introduced by \citet{Hartwig15b}. Our model produces self-consistent Pop III star formation histories and SN rates and accounts in an approximate fashion for the effects of radiative and chemical feedback. To check for consistency, we compare our PI SN rates with those from the FiBY simulations \citep{FiBY1}. Here, we briefly summarise the main features of the code. The modelling of star formation and feedback (Sections \ref{sect:SF} and \ref{sect:feedback}) is described in more detail in \citet{Hartwig15b, Hartwig2016a}. The merger trees (Section \ref{sect:trees}) are constructed with an algorithm from \citet{Parkinson2008} which is a modified version of \textsc{galform} \citep{galform}.

\subsection{Modified and unmodified EPS formalism}
\label{sect:trees}
Our merger tree algorithm is based on the extended Press Schechter (EPS) formalism first developed by \citet{Bond1991} and \citet{LaceyCole1993}, which produces the conditional halo mass function $\D N / \D M_1 (M_2, z, \Delta z$). This function is the mean number of haloes with masses in the range $M_1 \rightarrow M_1+\D M_1$ into which a halo with mass $M_2$ splits during the redshift interval $\Delta z$ at redshift $z$. Note that the merger tree is constructed backwards in time, starting with the most massive halo at $z=1$ and then reconstructing its assembly history by stepping backwards in time, to successively higher redshifts. The conditional halo mass function is used to create a probabilistic merger history of a dark matter halo.

The EPS formalism is known to underpredict the number of the most massive halo progenitors \citep[see e.g.][]{galform}. To remedy this deficiency \citet{Parkinson2008} modify the conditional halo mass function by
\begin{equation}
\frac{\D N}{\D M_1}(M_2, z, \Delta z) \rightarrow \frac{\D N}{\D M_1}(M_2, z, \Delta z) A \left(\frac{\sigma_1}{\sigma_2}\right)^{B}\left(\frac{\delta_2}{\sigma_2}\right)^{C},
\label{eq:eps_mod}
\end{equation}
where $\sigma_1, \sigma_2$ are the mean cosmic density variations on scales that correspond to $M_1$ and $M_2$ and $\delta_2$ is the critical overdensity for spherical collapse at redshift $z+\Delta z$. $A, B$ and $C$ are numerical factors, which are set to $A=0.57, B=0.38, C=-0.01$ to create merger trees with mass assembly histories consistent those in the Millennium simulation \citep{Millennium}. While we do not adopt the cosmological parameters used in the Millennium simulation ($H_0= 73\,\mathrm{km}\,\mathrm{s}^{-1}\mathrm{Mpc}^{-1},\ \Omega_m=0.25,\ \Omega_b=0.045,\ \Omega_{\Lambda}=0.75,\ \sigma_8=0.9,\ n_s=1$ \citealt{Millennium}), \citet{Parkinson2008} argue that their modifications are valid for a wide range of cosmological parameters. This is supported by the halo mass functions (HMFs) we produce in Section \ref{sect:repr}. By using WMAP7 cosmological parameters in a test run \citep[][except $\tau_e$]{wmap7year} we verify that the our SN rates are not very sensitive to cosmological parameters other than $\tau_e$.

\citet{Sasaki2014} showed that the abundances of haloes with masses $M \lesssim 10^8\,\Ms$ at high redshifts ($z\geq 15$) are reasonably well predicted by the Press-Schechter (PS) or Sheth-Mo-Tormen (SMT) halo mass functions \citep{PressSchechter, ShethMoTormen}. Even though the Parkinson algorithm has only been tested at relatively low redshifts ($z\leq 9$) we will demonstrate later that with the method described in Section \ref{sect:repr} we produce HMFs similar to PS or SMT HMFs at all relevant redshifts.

\subsection{Pop III star formation}
\label{sect:SF}
Star formation and feedback in our model was first implemented in \citet{Hartwig15b} and later improved in \citet{Hartwig2016a}. The model distinguishes between Pop III and Pop I/II star formation. As it is the subject of our study, Pop III SF is modelled in much more detail, while Pop I/II SF is only included as a background radiation field. We extrapolate the Pop I/II SF history from \citet{BehrooziSilk2015} and add its contribution to the ionising and Lyman-Werner (LW) backgrounds. 

In our fiducial model we assume that Pop III stars form only in metal free haloes, but we will also test a critical metallicity of $Z_\mathrm{crit}= 10^{-3.5}\, Z_\odot$. Metal-free or metal-poor haloes collapse as soon as H$_2$ cooling becomes sufficiently effective, i.e. as soon as the gas temperature exceeds a critical value $T_\mathrm{crit}$ \citep{Tegmark97}. Assuming the gas temperature is the virial temperature, the critical dark matter halo mass for Pop III star formation is taken as
\begin{equation}
  M_\mathrm{crit} = 1.0\times 10^6\,\Ms \left(\frac{T_\mathrm{crit}}{10^3\mathrm{K}}\right)^{3/2}\left(\frac{1+z}{10}\right)^{-3/2}.
\label{eq:m_cool_h}
\end{equation}
In our fiducial model we set $T_\mathrm{crit} = 2200\,\mathrm{K}$, e.g. as found in cosmological simulations by \citet{hum12}. To see how sensitive our model is to this parameter we also consider $T_\mathrm{crit} = 1100\,\mathrm{K}$ and $T_\mathrm{crit} = 4400\,\mathrm{K}$. Varying $T_\mathrm{crit}$ can also mimic how, for example, magnetic fields \citep{Schleicher09} or supersonic baryon streaming \citep{Greif2011a, Stacy2011, Maio2011} affect at what mass haloes collapse and how long the collapse takes.

We populate haloes with Pop III stars with masses randomly drawn from the IMF defined below until they exceed a total mass of
\begin{equation}
M_\mathrm{PopIII} = M_\mathrm{gas} f_* \eta,
\label{eq:SF_eff}
\end{equation}
where $M_\mathrm{gas}$ is the total baryon mass of the halo, $\eta$ is a star formation efficiency parameter and $f_*$ accounts for suppression of SF by LW feedback, as we will discuss below in Section \ref{sect:feedback}. We assume that the ratio of baryon mass to dark matter halo mass is equal to ratio of the corresponding cosmological densities $\Omega_b/(\Omega_m-\Omega_b)$. The star formation efficiency parameter $\eta$ is kept constant, but $f_*$ depends on the intensity of the LW background and on the mass of the collapsing haloes. Thus the overall Pop III star formation efficiency varies with redshift. For every individual run we calibrate $\eta$ so that our simulation produces a cosmological ionisation history that is consistent with observations, i.e. that reproduces the Thomson scattering optical depth $\tau_e$. From 2014 to 2016 $\tau_e$ has been revised from $0.092 \pm 0.013$ \citep{Planck2013} to $0.066 \pm 0.016$ \citep{Planck2015} and then to $0.054\pm0.013$ \citep{Planck2016}.  These changes have been shown to have significant implications for models of early star formation \citep{Visbal2015}. We consider all three values of $\tau_e$ but adopt 0.066 in our fiducial model. To compute the ionisation history we assume ionising photon escape fractions of $f_\mathrm{esc, Pop II}=0.1$ for Pop I/II stars and $f_\mathrm{esc, Pop III}=0.5$ for Pop III stars.

To create Pop III stars in our fiducial model we assume a logarithmically flat IMF with a lower limit $M_\mathrm{min} =1\,\Ms$ \citep{Greif11b}:
\begin{equation}
\frac{\D N}{\D \log M} =
  \begin{cases} 
      \hfill \text{const.}    \hfill & \text{if $M_\mathrm{min}< M <M_\mathrm{max}$} \\
      \hfill 0 \hfill & \text{otherwise}.
  \end{cases}
\label{eq:imf}
\end{equation}
We consider a range of values for the upper limit, $M_\mathrm{max}$. Because low-mass Pop III stars do not contribute significantly to reionisation and we calibrate the SF efficiency parameter $\eta$ with $\tau_e$, the total number of higher mass stars is essentially unaffected by the lower IMF limit, as long as it is of the order of a few solar masses or less.

\subsection{Feedback}
\label{sect:feedback}
Once Pop III stars form they affect subsequent star formation in several ways. The LW radiation they emit gradually builds up a background that photodissociates molecular hydrogen and impedes cooling and SF in haloes that have not yet formed stars. This effect reduces the total mass that goes into primordial star formation by a factor $f_*$. Following \citet{Machacek2001}, $f_*$ can be approximated as
\begin{equation}
f_* = 0.06 \ln \left(\frac{M_\mathrm{Halo}/\Ms}{1.25\times 10^5+8.7\times 10^5 F_\mathrm{LW}^{0.47}} \right),
\label{eq:LW_feedback}
\end{equation}
where $F_\mathrm{LW}$ is the LW flux in units of $10^{-21}$ erg s$^{-1}\,\mathrm{cm}^{-2}\,\mathrm{Hz}^{-1}$. If $f_*$ calculated from Equation \eqref{eq:LW_feedback} would be zero or smaller no gas can collapse and SF is delayed until $f_*>0$. To calculate the LW flux we use the \citet{Schaerer2002} stellar evolution models and LW escape fractions from \citet{Schauer2015}.

When the first SNe occur they enrich their environments with metals. Since SN remnants can grow to be much larger than their host haloes, Pop III SNe can contaminate other haloes with metals, causing them to form new stars with much lower characteristic masses. However, dense gas in collapsing minihaloes does not easily mix with metals from nearby SNe \citep{Cen2008}. We assume therefore that external enrichment happens by mixing the gas with metals before the halo reaches $M_\mathrm{crit}$.

Assuming a random spatial distribution of SNe we calculate the chances for the gas of a currently collapsing halo to overlap with at least one SN remnant. If the halo is being enriched we randomly select a SN remnant with a probability corresponding to its volume \citep[see][Section 2.4.1]{Hartwig15a}. We then compute the mass of the metals that enrich the halo by multiplying the metal surface density of the supernova remnant by the halo cross-section, computed at the virial radius. For the metal surface density we assume that all the metals ejected by the supernova are uniformly distributed over a thin outer shell of the remnant \citep{bsmith15, Ritter16}. This procedure allows us to estimate the metallicity of the externally enriched haloes in a statistical sense.

Dynamical heating during mergers can also prevent the gas from cooling. If the current mass growth rate of a halo is too large, i.e. while
\begin{equation}
 \frac{\D M}{\D z} \geq 3.3 \times 10^6\, \Ms \left(\frac{M}{10^6\,\Ms} \right)^{3.9},
\end{equation}
we assume that Pop III SF is suppressed \citep{Yoshida2003}.

To this previously existing feedback we add feedback due to reionisation: If a minihalo is located in an ionised region, the gas temperature ($T \approx 10^4\,\mathrm{K}$) is much higher than its virial temperature \citep[e.g.][]{GloverReview}. Such a halo does not collapse until it reaches a much higher mass, by which time it has most likely been enriched with metals by one of its progenitors or by external sources. Therefore, when a pristine halo reaches the critical mass for collapse, we prevent Pop III SF with a probability that is equal to the current ionisation fraction of the Universe.

\subsection{Creating cosmologically representative data}
\label{sect:repr}
The SF model we use originally was designed to simulate Pop III SF in the progenitors of a single low-redshift massive halo. Running a cosmologically representative set of these low redshift haloes would require us to simulate a very large number of low mass merger trees for every high mass merger tree. Instead, we simulate a reduced sample of merger trees and weigh the results according to the number density of initial halos of similar mass.

Starting with low redshift haloes at $z=1$ we set up the code to run for $N_\mathrm{HMF}=300$ different halo masses $M_i,\ 1 \leq i \leq N_\mathrm{HMF}$ between $M_\mathrm{Halo, min} = 5 \times 10^8\, \Ms$ and $M_\mathrm{Halo, max} = 2 \times 10^{13}\, \Ms$. We distribute the mass bins according to a power-law with exponent $\alpha$. Several random realisations are computed for each mass. We assume that the merger tree with mass $M_i$ is representative for merger trees with masses $M_{i,\mathrm{low}}\leq M \leq M_{i,\mathrm{up}}$, where $M_{i,\mathrm{up/low}}=0.5(M_i+M_{i\pm1})$. To get cosmologically representative densities of e.g. SNe or the number of ionising photons, the results from each tree are weighted according to the comoving number density of haloes in the corresponding mass bin and added up. This weighting factor is
\begin{equation}
 w_i = \frac{1}{M_i}\int_{M_{i,\mathrm{low}}}^{M_{i,\mathrm{up}}} M \frac{\D N}{\D M} \D M,
 \label{eq:weights}
\end{equation}
where $\D N /\D M$ is the SMT HMF calculated with \textsc{HMFcalc} \citep{HMFcalc}. These weighting factors have the convenient property of preserving the total matter density when we change $N_\mathrm{HMF}$ or $\alpha$. Thus, as long as the HMF is sampled reasonably well, neither $N_\mathrm{HMF}$ nor $\alpha$ have an impact on our SN rates or the computed ionisation histories. However, changing $\alpha$ allows us to tune what fraction of the computational time is spent on low-mass or high-mass merger trees.

The lower limit of the initial halo mass distribution is chosen such that each merger tree contains at least a few Pop III star-forming haloes and thus produces a representative star formation history. A merger tree with a much lower initial halo mass would produce only a single Pop III forming halo at low redshifts in the absence of feedback from previous Pop III SF. For these low mass merger trees the major feedback component would come from outside the tree and could not be modelled with our code. Our SN rate distributions are not sensitive to increasing $M_\mathrm{Halo, min}$ by a factor of a few and change only by a few percent. The upper limit is chosen so that haloes become too rare to contribute significantly to a cosmologically representative sample. We choose a power law exponent of $\alpha = -1.3$ for the distribution of initial halo masses.

\subsection{Calculating SNe rates}
We count the number of SNe occurring for each tree and average over the different realisations. Weighting with the abundance factors $w$ defined in Section \ref{sect:repr} yields the comoving SN rate density $\D N/(\D V\D t)$. Following \citet{hum12}, we calculate the redshift distribution of the observable SN rate per unit solid angle as
 \begin{equation}
 \begin{split}
 \frac{\D N}{\D t_\mathrm{obs}\D z\D\Omega} &= \frac{\D N}{\D t_\mathrm{obs} \D V} \frac{\D V}{\D z \D\Omega}\\
 & = \frac{1}{1+z} \frac{\D N}{\D t \D V} {r(z)}^2 \frac{\D r}{\D z},
 \end{split}
  \label{eq:z_distr}
 \end{equation}
 where the physical time $\D t$ is converted to observer time with $\D t_\mathrm{obs} = (1+z) \D t$ and $r(z)$ is the comoving distance to the redshift of the step $\D z$.
 
\subsection{A closer look at halo mass functions}
\begin{figure}
 \includegraphics[width=\linewidth]{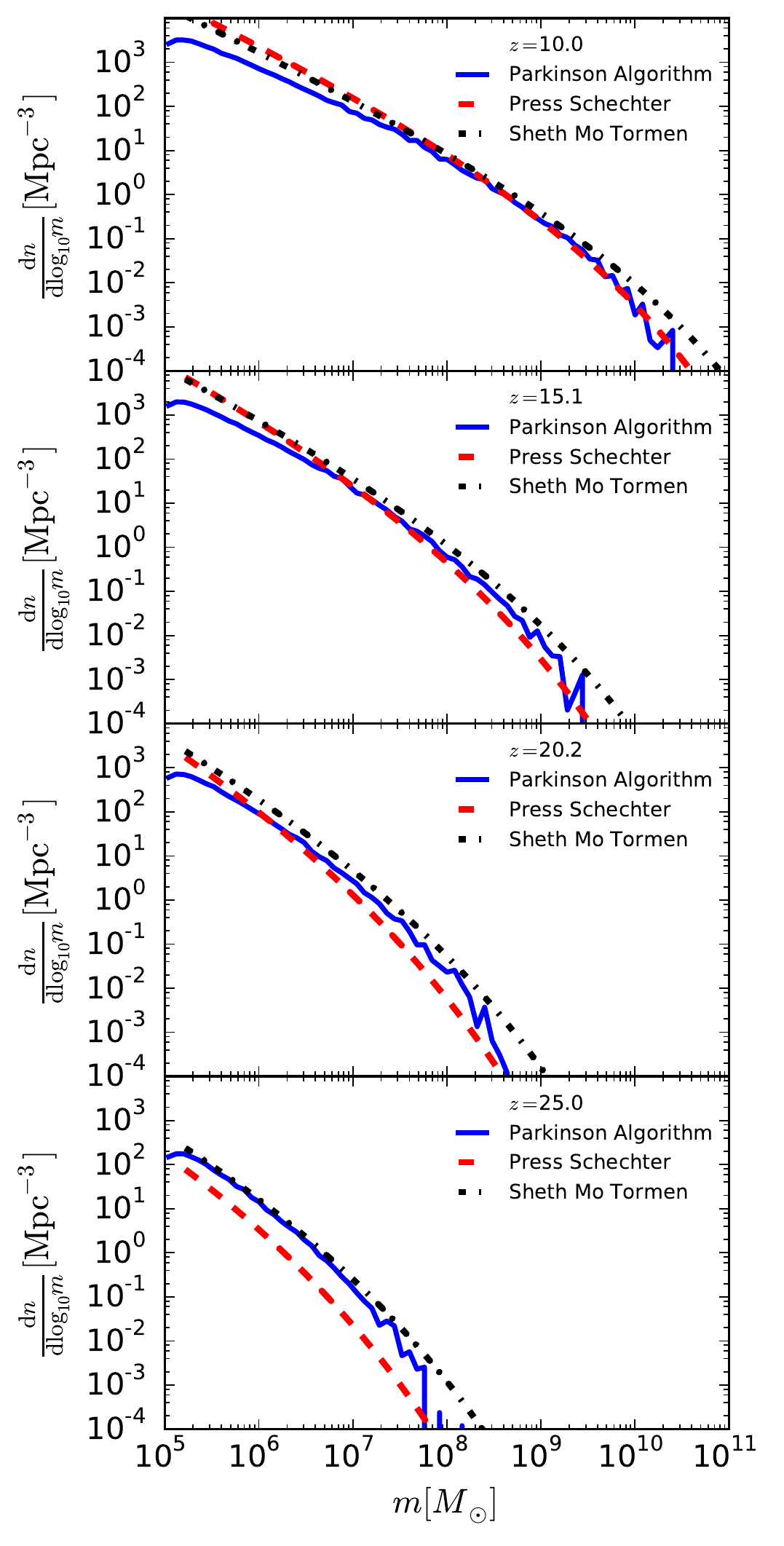}
 \caption{\label{fig:hmf}Halo mass functions from the Parkinson algorithm with representative initial halo sampling (equation \ref{eq:weights}) for $z\sim 10 - 25$. The halo mass function roughly agrees with the PS and SMT mass functions at all redshifts.  At lower redshifts a fraction of the lowest mass haloes are excluded because of the lower mass limit for the initial halo mass function.}
\end{figure}

The accuracy of the Parkinson code has primarily been tested and confirmed for $z\lesssim 10$ \citep{Parkinson2008, Jiang2014}. To calculate SN rates it is especially important that the algorithm also predicts reasonable numbers of minihaloes at higher redshifts. As shown in Fig. \ref{fig:hmf}, with our cosmologically representative initial halo sampling the Parkinson algorithm reproduces PS \citep{PressSchechter} and SMT \citep{ShethMoTormen} HMFs, even at high redshifts. At redshift 25 where PS and SMT mass functions deviate significantly, our models follow the SMT HMF.  With high-resolution cosmological dark matter simulations, \citet{Sasaki2014} showed that these mass functions and especially the SMT HMFs predict the right minihalo densities at $z\geq 15$.

\subsection{List of models}
\label{sect:list}
\begin{table}
 \begin{tabular}{lll}
 Label & $\eta$ &Description\\\hline
 Fiducial & 0.35 & Fiducial model\\
 T4400 & 0.13 & Critical temperature $T_\mathrm{crit} = 4400\, \mathrm{K}$\\
 T1100 & 0.7 & Critical temperature $T_\mathrm{crit} = 1100\, \mathrm{K}$\\
 FiBY\_rep & 0.01 & FiBY reproduction run\\
 &&(see Section \ref{sect:fiby})\\
 old $\tau_e$ & 1.1 & adjusted to match $\tau_e = 0.096$ \\
 3 keV WDM & 0.22 & 3$\,$keV warm dark matter model\\
 5 keV WDM & 0.15 & 5$\,$keV warm dark matter model\\
 $Z_\mathrm{crit}$ & 0.12 & Using $Z_\mathrm{crit}=10^{-3.5}\, Z_\odot$ as \\&&critical value for Pop III SF\\
 IMF\_300 & 0.34 & log flat IMF with upper limit 300$\,\Ms$\\
 IMF\_170 & 0.38 & log flat IMF with upper limit 170$\,\Ms$\\
 IMF\_60 & 0.6 & log flat IMF with upper limit 60$\,\Ms$
 \end{tabular}
\caption{\label{tab:runs}Pop III star formation models. A large deviation to the $\eta$ in the fiducial model indicates that changes may have a large influence on the physics, but as we normalize the star formation history such that it produces a reasonable ionisation history, the impact on the final SN rates can still be negligible.}
\end{table}
The Pop III star formation models in our study are listed in Table \ref{tab:runs}. Some additional models were performed as stability tests: two had different lower limits for the IMF (0.1 and 3.0\,\Ms), one had a different lower limit on the HMF (2.0$\times 10^{9}\,\Ms$) and one used WMAP7 cosmological parameters \citep[][except $\tau_e$]{wmap7year}, i.e. with $H_0= 70.4\,\mathrm{km}\,\mathrm{s}^{-1}\mathrm{Mpc}^{-1},\ \Omega_m=0.272,\ \Omega_b=0.0455,\ \Omega_{\Lambda}=0.728,\ \sigma_8=0.811,\ n_s=0.976$. None exhibited noticeable departures from the SN rates in the fiducial run. We also performed a run with the most recent $\tau_e = 0.054$ in which the total star formation rates (SFR) and SN rates fell by 30 per cent but the redshift distribution of SF and SNe were not affected.

\section{Results}
\label{sect:res}
\subsection{Our fiducial model and its sensitivity to the choice of IMF}
\begin{figure}
 \includegraphics[width=\linewidth]{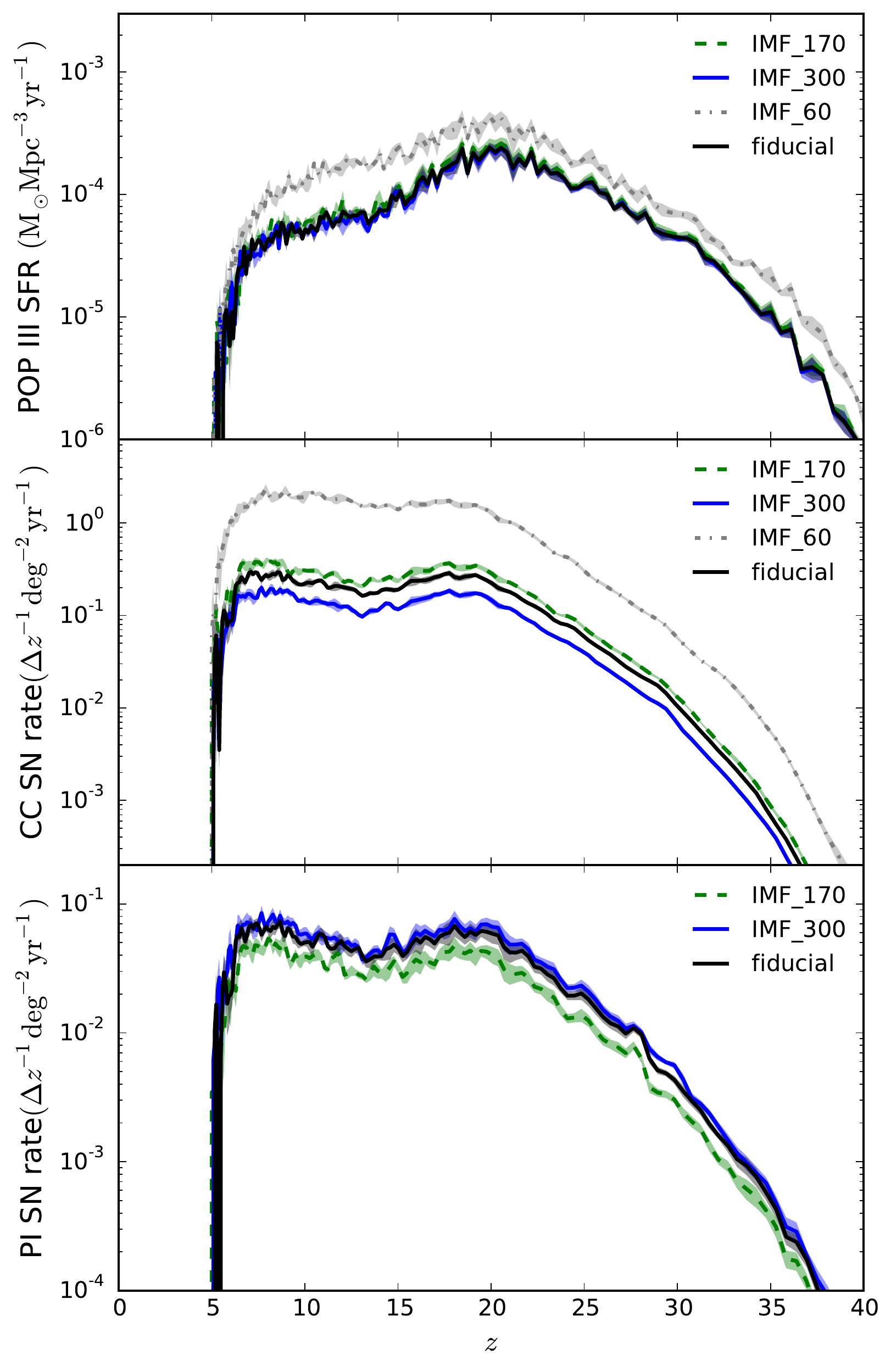}
 \caption{\label{fig:fid}Pop III SFR densities (upper panel), CC SN rates (centre panel) and PI SN rates (lower panel) for our fiducial model and the alternative IMFs we test. SFRs are presented as stellar mass per comoving Mpc$^3$ and year proper time, the SN rates are measured in number per year observer time, square degree and unit redshift. Because we calibrate the star formation efficiency parameter $\eta$ with the optical depth $\tau_e$ the total mass in massive stars remains roughly constant with IMF, which leads to similar PI SN rates for most of the IMFs. The exception is the IMF\_60 model, which by construction does not produce any PI SNe, but which consequently produces a large number of CC SNe. The statistical noise in the IMF models is similar because the merger trees are based on the same random number seed.}
\end{figure}

In Fig. \ref{fig:fid} we show Pop III star formation rate densities and PI and CC SN rates for our fiducial model. The total PI SN rate is 1.0 SN yr$^{-1}$ deg$^{-2}$, which corresponds to one PI SN yr$^{-1}$ per 3 \emph{WFIRST} fields of view (FoV)\footnote{Assuming a 0.4$^\circ\times 0.8^\circ$ FoV taken from \emph{WFIRST} reference cycle 6: \url{http://wfirst.gsfc.nasa.gov/science/Inst_Ref_Info_Cycle6.html}} or per 370 \emph{JWST} FoV\footnote{Assuming a 2.2$^\prime\times 4.4^\prime$ FoV for the \emph{JWST} NIRCam \citep{jwst06}}. The total CC SN rate for the fiducial model is 4.2 SNe yr$^{-1}$ deg$^{-2}$.

SFRs and SN rates have a different dependence on redshift because the SFRs are measured per year proper time while the SN rates are measured per year observer time. We do not find Pop III SF at $z < 5$ due to chemical and ionisation feedback. The SFRs peak at $z\approx 20$ and decrease towards lower redshifts. The extrapolated Pop II SFRs surpass the Pop III SFRs around $z=20$ where the Pop III SFRs reach their peak. In contrast to the SFRs, the SN rates are roughly flat from $5 < z < 20$, which is caused by projecting the SN rate densities on a unit solid angle of the sky with Equation \eqref{eq:z_distr}, i.e. by  accounting for time dilation and the redshift dependence of the angular diameter distance. The abrupt drop in SFRs and SN rates at $z>20$ is due to the rapid change in the abundance of minihaloes that could form stars. The shaded regions indicate the statistical fluctuations between the different random realisations of the merger trees. In Fig. \ref{fig:fid} and \ref{fig:misc} there is a small systematic bump in the PI SN rates at $z\sim 27$ which is caused by our time steps being redshift dependent. There is a transition in how many time steps the stars survive. So PI SNe from stars born in two different time steps explode in the same time step. To avoid this issue, we would have to properly resolve the lifetimes of all PI SN stars at all times, which would require time steps of roughly 10$^5$ years over the whole redshift range. For our model using these small time steps even at low redshifts is not viable. However, the effect on the overall PI SN rate distribution is small.

We also examine the effect the upper mass limit for the Pop III IMF has on the SN rate, as described in Section \ref{sect:SF}. As long as the IMF has significant overlap with the PI SN mass range, the PI SN rate is not strongly affected by the upper IMF limit. Our IMF\_300 and IMF\_170 models yield total PI SN rates of 1.1 and 0.7 events yr$^{-1}$ deg$^{-2}$, which are both very close to our fiducial model. Of course, the PI SN rates can be made arbitrarily low by reducing or completely removing the overlap between the IMF and the PI SN mass range. This is demonstrated in our IMF\_60 model, which does not produce PI SNe but has larger SFRs by a factor of a few. The increased SFRs are caused by the higher star formation efficiency needed to produce the same optical depth with less massive stars. In this model the total CC SN rate increases to 29 SNe yr$^{-1}$ deg$^{-2}$, almost an order of magnitude higher than in our fiducial model.

\subsection{Comparison to previous estimates}
\label{sect:fiby}

To verify that our method produces reasonable SN rates we compare them to those from the FiBY simulations \citep{FiBY1}. For this we use their Salpeter IMF with the mass range 21 -- 500 \Ms. However there are some key differences with our model. We impose an upper limit on our HMF sampling of $9\times10^{10}\,\Ms$ to account for their limited simulated volume. At $z=1$ we expect only one halo that is $9\times10^{10}\,\Ms$ or higher in their $64\, \mathrm{Mpc}^3$ simulation volume. Because one generally cannot simulate `half a halo', the upper end of the HMF can be significantly biased by the finite volume. Also, we modify our cooling criterion to allow SF only in haloes more massive than $2\times 10^7\,\Ms$. This mass is roughly consistent with the typical halo mass for Pop III SF in the FiBY simulations \citep{Paardekooper2013}. However, there are some effects for which we do not account that could have similar impacts on the SN rates, as changing the critical collapse mass by a factor of a few. First, there is no direct correspondence between our SF efficiency parameter $\eta$ and their criteria for SF, so it is not clear if $\eta$ should be varied with redshift or halo mass to arrive at similar numbers of Pop III stars per halo. Also, it is unclear how strongly their HMF is biased by their finite simulation volume and by statistical scatter in the number of rare haloes.

\begin{figure}
 \includegraphics[width=\linewidth]{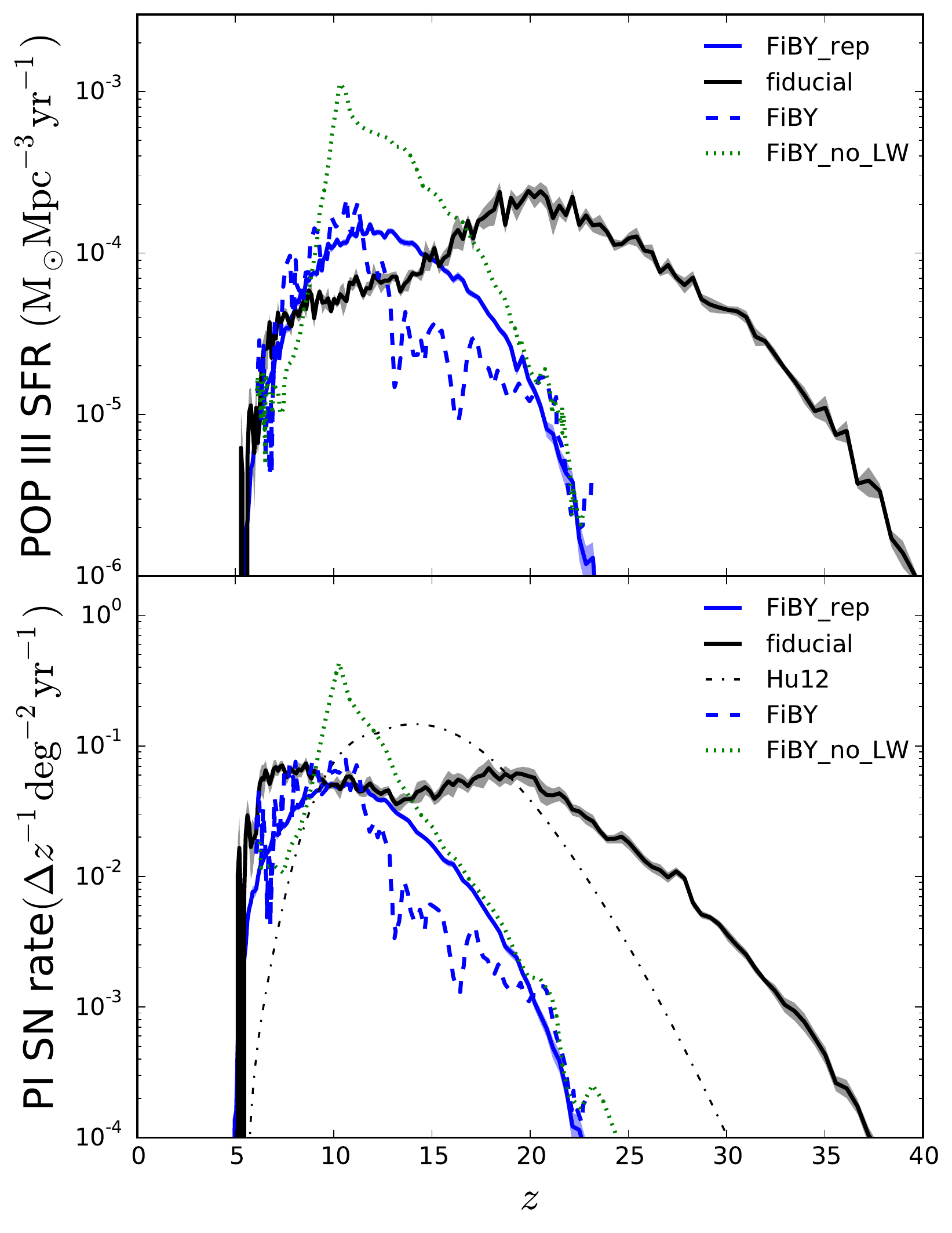}
 \caption{\label{fig:fiby_rep}Pop III SFR densities (upper panel) and PI SN rates (lower panel) for the FiBY reproduction run. Increasing the critical mass for collapse to the level of what is observed in FiBY yields similar PI SN rates. At some redshifts our rates are closer to their no--LW control run because our LW feedback has less effect on the minimum halo mass for SF in our run. Our fiducial model predicts larger PI SN rates at high redshifts. We also plot PI SN rates from \citet{hum12} for reference.}
\end{figure}

Figure \ref{fig:fiby_rep} shows that we nevertheless reproduce the PI SN rates predicted by FiBY relatively well. Our SFRs and PI SN rates lie between their fiducial rates and control run rates without LW feedback, because the LW feedback we use \citep[from][]{Machacek2001} has little impact on haloes above $10^7 \Ms$. This agreement with simulations suggests that our semi-analytical model captures much of the complexity of Pop III SF, at least in a statistical sense.

\citet{hum12} also estimated global PI SN rates and detection rates for \emph{JWST}. Based on the PS formalism, they first calculate the rate at which haloes reach the critical mass. They then assume that in such a halo a single 100 \Ms\ Pop III star forms and explodes as a PI SN. They also include radiative and chemical feedback. We compare our rates to those in their enhanced SF(ESF) scenario, which are intermediate between their conservative feedback and no feedback models. In the ESF model, Pop III star formation is suppressed by chemical feedback and enhanced by radiative feedback. Because haloes can grow to larger masses before collapsing in the presence of a strong LW radiation field, more stars form in each halo and the number of SNe per halo rises. Our fiducial rates are close to theirs overall but are greater at the highest redshifts. Our HMFs are closer to the SMT mass function, whereas their calculation is based on the PS formalism, so the number of star forming minihaloes we predict at $z\gtrsim 30$ is higher. The remaining difference can be attributed to the IMF they assume. Their integrated PI SN rate is about 1 yr$^{-1}$ deg$^{-2}$, very similar to our fiducial model. \citet{wa05} predict an integrated PI SN rate of 0.34 yr$^{-1}$ deg$^{-2}$, which is lower than our fiducial rate and the \citet{hum12} rate because they assume only one PI SN per Pop III star forming halo.

At present we do not reproduce PI SN rates from the Renaissance simulations. Pop I/II SFRs exceed Pop III rates at very early times in some of these models \citep[at $z\sim 27$ in their rare peak model;][]{xu13}. The total mass in Pop II stars exceeds the Pop III mass 20\,Myr after the first Pop III star is born. These early high Pop II SFRs are inconsistent with the Pop II SF history we assume and also with that in the FiBY simulations \citep{FiBY1}. For example at redshift $z=15$ our extrapolated Pop II SFRs and the FiBY SFRs are both on the level of a few times $10^{-4}\,\Ms$yr$^{-1}$pc$^{-3}$, while the Renaissance simulations have Pop II SFRs that are an order of magnitude higher for the medium density run and $0.1\,\Ms$yr$^{-1}$pc$^{-3}$ for the rare peak run \citep{XuNorm}. The transition from Pop III to Pop II SF in large-scale cosmological simulations is sensitive to the criteria adopted for the formation of Pop II stars and the resolution with which the models can capture metal mixing with nearby haloes.  At the mass resolution of the FiBY and Renaissance simulations the latter is never achieved. This may have led to overmixing and a premature transition from Pop III to Pop II SF in some of the Renaissance models. Future simulations with the requisite resolution will be needed to definitely state when Pop II SFRs surpass Pop III rates globally.

\subsection{Warm dark matter and SNe as probes of structure formation}
While the $\Lambda$ cold dark matter ($\Lambda$CDM) model has been very successful in explaining the large scale structure of the Universe, there are multiple problems at smaller scales \citep{DOnghia2004, Kroupa12, Pawlowski15}. $\Lambda$CDM has not been able to reproduce the missing dark matter substructure in galactic haloes and the small number of satellite galaxies of the Milky Way. Warm dark matter (WDM) is a possible solution to this 'small scale crisis' of the $\Lambda$CDM model. In this picture, dark matter consists of roughly keV-mass particles, which results in the suppression of structure formation on small scales. WDM can be implemented in our model as a modified dark matter power spectrum and the corresponding changes to the HMF. To compute WDM power spectra and HMFs we use the \textsc{HMFcalc} web tool\footnote{\url{http://hmf.icrar.org/}} \citep[][see Section 2.3]{HMFcalc}. We focus on WDM that is a thermal relic with a particle mass of 3\,keV, which lies at the strong impact limit of the observational constraints \citep{Viel2013}. We also test a weaker 5\,keV WDM model. We do not discuss CC SN rates separately in this and the next Section because we keep the Pop III IMF used in the fiducial model and thus the ratio of PI SNe to CC SNe stays fixed.

\begin{figure}
\includegraphics[width=\linewidth]{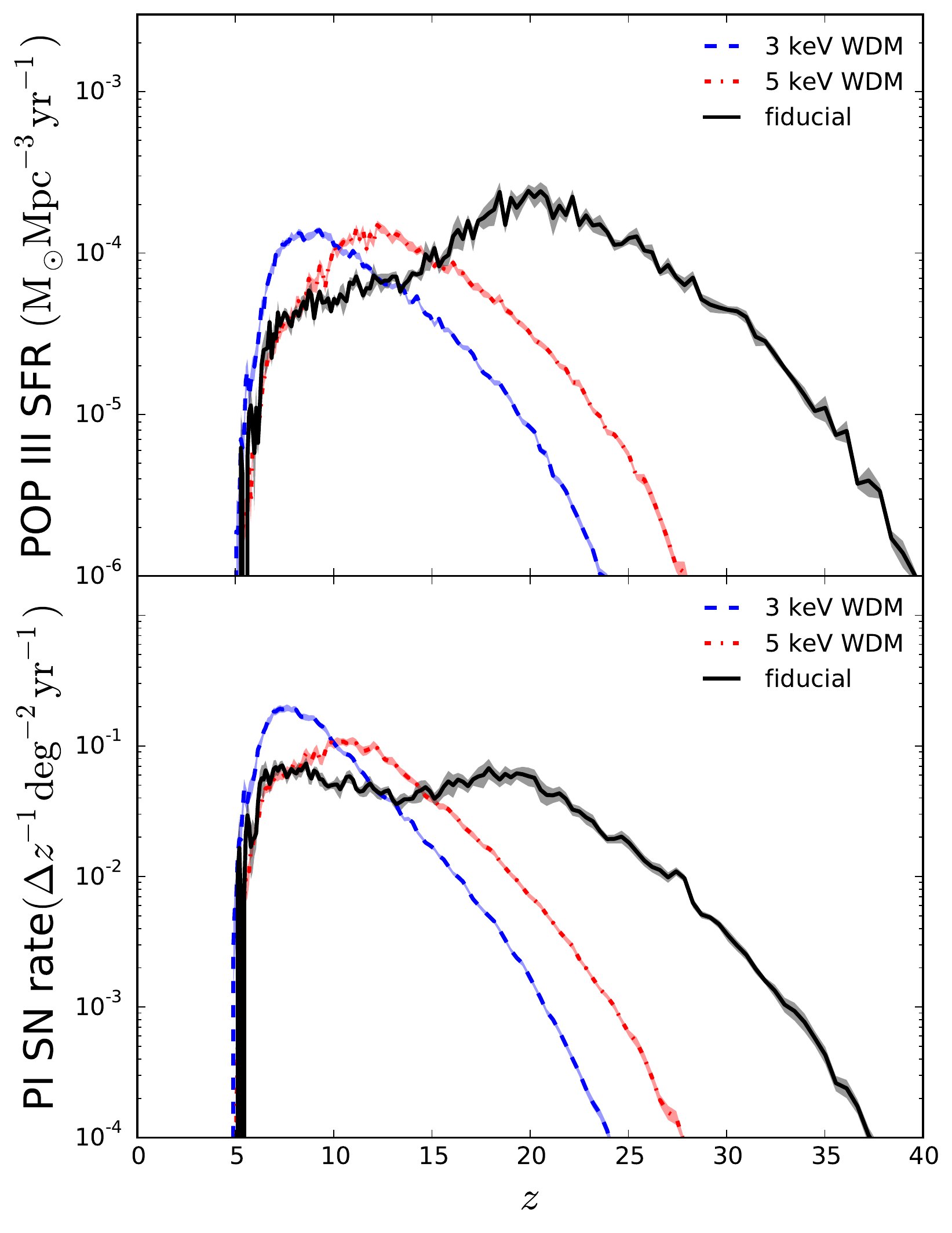}
\caption{\label{fig:wdm_rates} Fiducial and WDM Pop III SFR densities (upper panel) and PI SN rates (lower panel). The first PI SNe are significantly delayed in a 3 keV WDM Universe and somewhat delayed with 5 keV WDM particles. Thus, very high redshift PI SN detections could place a lower limit on the WDM particle mass.}
\end{figure}

In Fig.~\ref{fig:wdm_rates} it is clear that that WDM mainly suppresses early Pop III SF and SN rates. \citet{Maio2015} find a similar effect, although at lower redshift because Pop III SF occurs in much more massive haloes in their simulations. Pop III SF begins at very high redshifts in a pristine Universe with no feedback from previous SF. It therefore only depends on when the first structures appear that are massive enough to host SF. That is, it is only sensitive to the cosmological model and the critical halo mass for Pop III SF. For these reasons, the SN rate curve is unique to each cosmology (although the total SN rate can be influenced by other physical parameters -- see Section \ref{sect:caveat}).  This feature makes high redshift SNe a suitable probe of structure formation and the nature of dark matter.
 
\begin{figure}
\includegraphics[width=\linewidth]{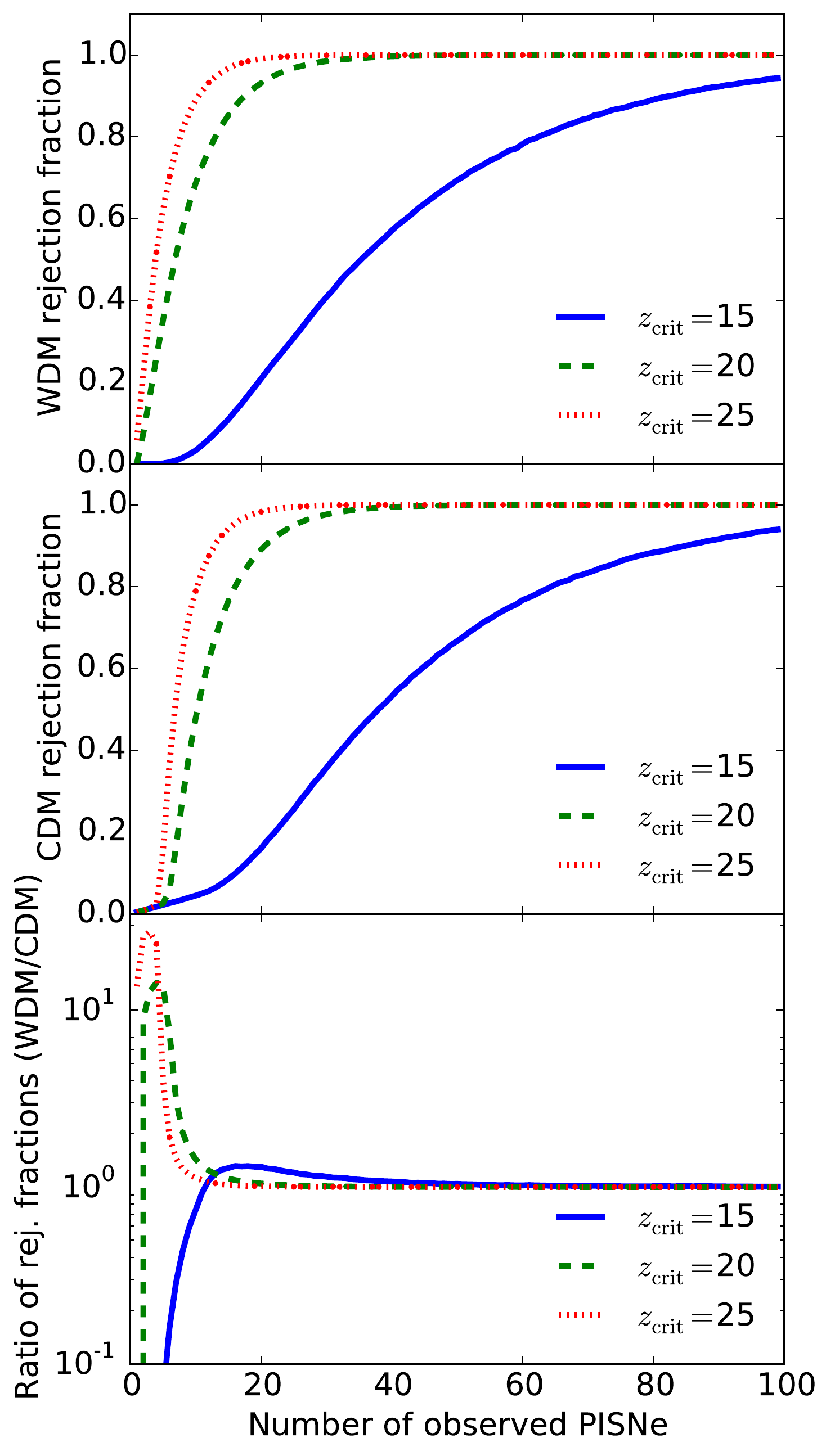}
\caption{Upper panel: Rejection probability of the $3$\,keV WDM model to $99\,$ per cent confidence in a CDM cosmology, as a function of sample size for different maximum observable redshifts. Middle panel: same but with WDM and CDM swapped. For simplicity we assume that all SNe are visible for the same amount of time up to a certain redshift. Lower panel: WDM rejection probability divided by CDM probability. Even small numbers of PI SN observation could be enough to reject certain WDM models. The typical WDM case -- only observations at lower redshifts -- could also be produced at random in a CDM model. Thus, for observations reaching $z=20$ or $z=25$ and small sample sizes, the WDM rejection fractions rise faster than the CDM rejection fractions.}
\label{fig:obs_prob}
\end{figure}

In Fig.~\ref{fig:obs_prob} we plot the probability of rejecting either CDM or $3\,$keV WDM under the assumption that one of the two models is correct as a function of the number of observed SNe. SNe at higher redshift become fainter but are visible for longer times because of cosmic time dilation. For simplicity, we assume that these effects cancel up to a certain redshift; i.e, that PI SNe are directly observable for the same time up to a critical redshift $z_\mathrm{crit}$ and not above it. Realistic predictions of the observability of SNe with current or future telescopes strongly depend on instrumental details and survey strategies and will be examined in a separate study (Rydberg et al., in prep.). Under these assumptions, our SN rates truncated at $z_\mathrm{crit}$ represent the redshift distributions of observed SNe. We draw a sample of SNe from the fiducial CDM distribution and check with a likelihood ratio test whether the sample is still compatible with the WDM distribution function. We reject the WDM model if the likelihood ratio is less than 0.01. For a range of sample sizes and 10$^4$ randomly drawn samples for each sample size we compute the fraction of samples that allows us to reject WDM. We repeat the same analysis with the roles of CDM and WDM swapped to probe the probability of rejecting CDM in a Universe that is well-described by WDM.

Rejecting a WDM model with this method would be easier than rejecting CDM for two reasons. On the one hand, if the observations reach high enough redshifts, in CDM there could be PI SNe detections at redshifts where there are close to no SNe in WDM. Therefore a sample of two or three PI SNe can already be incompatible with WDM. The typical WDM observation -- only SNe at lower redshifts -- can occur at random in CDM if the sample size is small. This effect is reflected by the $z = 20$ and $z = 25$ rejection fractions rising later in the middle panel of Fig. \ref{fig:obs_prob}. For 4 PI SN detections with a maximal detection redshift of $z=25$ the fraction of samples that rebut WDM is 50 per cent while the CDM rejection fraction is only 2 per cent. To emphasise the difference between WDM and CDM rejection fractions we plot their ratio in the lower panel of Fig. \ref{fig:obs_prob}. For the lowest maximum redshift and small sample sizes the rejection fractions of CDM are higher because the distribution is more peaked. For larger sample sizes the rejection fractions for CDM and WDM are very similar. On the other hand, even if there are detections that follow our WDM distributions and are significantly different from our CDM distributions, this could be explained by delayed Pop III SF e.g. due to strong LW feedback or supersonic baryon streaming. The constraint that Pop III SNe can not form above a certain redshift in WDM is much more solid, because of the simple lack of dark matter structures they could form in. If the observations only reach $z\lesssim 15$ the required sample size for rejecting either of the models increases by a factor of a few. In this case our analysis is relying on parts of the distribution that are impacted by the details of feedback modelling. They are not deviating as strongly as the high redshift distributions. Thus constraints on structure formation, that are based on observations that can not reach beyond $z\approx 15$ are much weaker.

\subsection{Other parameter tests}
\begin{figure}
\includegraphics[width=\linewidth]{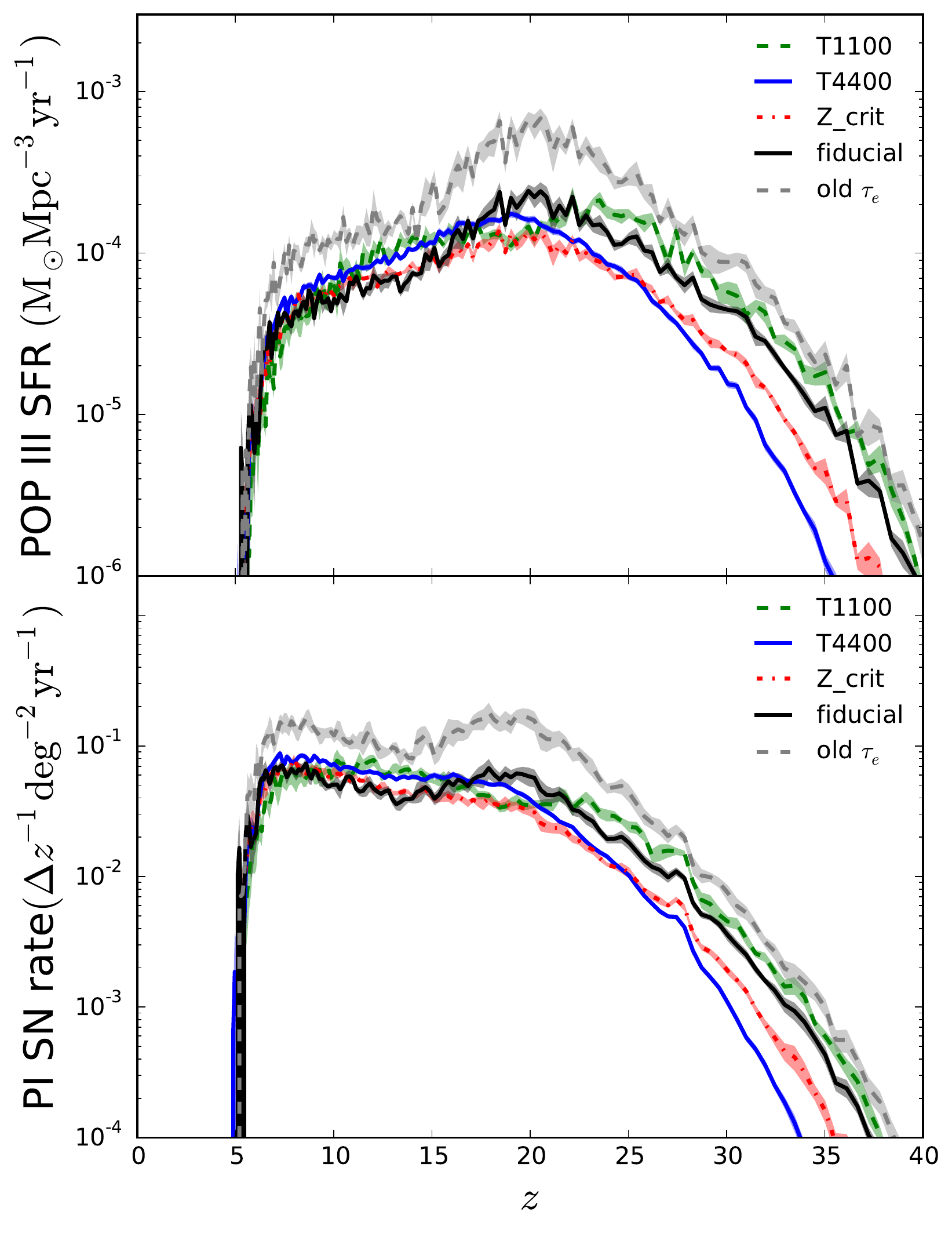}
\caption{Pop III SFR densities (upper panel) and PI SN rates (lower panel) for various parameter tests. Changing the critical temperature or the metallicity has only a minor effect on the PI SN rates. Manipulating the required ionising photon budget (i.e. using a larger $\tau_e$) has a much larger impact on the total rates. The effects on the distribution of PI SNe is small for all tested parameters.}
\label{fig:misc}
\end{figure}

In Fig.~\ref{fig:misc} we consider the effect of other parameters on the PI SN rates and the SFRs. Decreasing the critical halo mass for Pop III SF by a factor of $\sim 3$ (T1100 model) has nearly no impact on the rates, yielding a similar rate curve and a total of 1.1 PI SNe yr$^{-1}$ deg$^{-2}$. Increasing the critical mass by the same factor (T4400 model) delays the onset of Pop III SF by $\Delta z \sim 5$ but still leads to the same total rates as our fiducial model. In the Z\_crit run Pop III SF happens in roughly twice as many haloes. Still, because a lower star formation efficiency is required to fit the $\tau_e$ parameter, the SN rate curves retain their shape and the total PI SN rates only change to 0.8 SNe yr$^{-1}$ deg$^{-2}$. As long as the overall picture of Pop III SF does not change, e.g. by altering the critical halo mass for Pop III SF by more than an order of magnitude (as in Section \ref{sect:fiby}), these parameters have little impact on the PI SN rates.

On the other hand, changing the ionising photon budget has a strong impact on the rates. Calibrating to the old $\tau_e$ parameter increases the total PI SN rate from 1.0 to 2.5 SNe yr$^{-1}$ deg$^{-2}$. This is important, as it illustrates the variations that can be caused by changing our calibration. Still, the redshift distribution of the PI SN rates remains very similar to our fiducial model.

\section{Caveats}
\label{sect:caveat}

One issue with our model is, that it is difficult to constrain the Pop III SF efficiency as there are no direct observations. Calibrating via the Thomson scattering optical depth $\tau_e$ is one solution but also adds additional uncertainties. The Pop III SF efficiency is strongly dependent on and degenerate with:
\begin{itemize}
 \item the ionising photon escape fractions for Pop I/II and for Pop III,
 \item the assumed Pop II SF history,
 \item the Thomson scattering optical depth $\tau_e$.
\end{itemize}
A wide variety of results can be produced by appropriately adjusting these parameters. For example with a Pop II ionising photon escape fraction of 20 per cent no Pop III stars would be needed for reionisation. Our model uses the best available data in its treatment of these parameters \citep[see][]{Hartwig15b}, but improvements in our knowledge of e.g. high redshift Pop II SF will have a direct effect on the accuracy with which we can predict Pop III SN rates. Also, we neglected active galactic nuclei, which can significantly contribute to the ionising photon budget \citep{Volonteri2009, Madau15}. We find that several hundred to a few thousand solar masses of Pop III stars are formed in most haloes in our fiducial model. This is in the range of typical masses of Pop III clusters found in numerical simulations \citep[e.g.][]{Stacy16, hir13}. At low redshifts, when Pop III forming haloes are much more massive, we form up to ten times more mass in stars per halo than at high redshift. So, despite the uncertainty in $\eta$ we are in a reasonable range with the stellar mass per halo. A factor of a few in uncertainty remains.

A further important source of uncertainty in our model is the lack of spatial information for the haloes in our merger trees. Radiative and chemical feedback between haloes should be strongly correlated with the overall structure of the merger tree. This could be overcome in the future by applying our early SF models to spatially resolved merger trees imported from large high resolution cosmological dark matter simulations, such as the Caterpillar project \citep{Caterpillar}.

Because of the mass limits on the initial haloes we can probe, we underestimate the number of minihaloes at low redshift (Fig.~\ref{fig:hmf}). This might artificially reduce the number of SNe predicted at low redshifts. However, for the low mass merger trees we would also need to consider feedback from outside the merger tree. As we currently cannot do this, we are also overestimating the SN rates at low redshifts. We also expect the low redshift Pop III SN rates to be even further reduced by feedback by later generations of stars. While our models show at which redshift feedback starts to significantly affect the SN rates, the exact strength of feedback at low redshifts remains uncertain.

It must be kept in mind that it is still an open question whether Pop III PI SNe actually occur, i.e., whether the Pop III IMF and the PI SN mass range overlap. While some simulations suggest that at least some Pop III stars are in the mass range for PI SNe \citep[e.g.][]{yoha06,hir13}, there are others in which such stars do not seem to form \citep[e.g.][]{clark11,Stacy16}. Also, extending the IMF to much higher masses or changing its slope could significantly reduce the number of PI SNe without changing the total mass in very massive Pop III stars. However, several PI SN candidates have now been discovered in the Universe today \citep{gy09,cooke12} and multiple stars with masses well above the threshold for PI explosions have also been found in the Local Group \citep{R136}, so it is plausible that such events did occur in the high redshift Universe.

\section{Summary \& conclusion}
\label{sect:conc}

We have estimated cosmologically representative Pop III SN rates and SFRs with a semi-analytical merger tree model. We explored how these rates are affected by a range of poorly constrained parameters. We calibrate the Pop III SF efficiency such that our model reproduces the observed Thomson scattering optical depth $\tau_e$. Therefore, our SN rates are not sensitive to some of the other parameters that govern Pop III SF. Specifically we found that:

\begin{itemize}
 \item As long as there is significant overlap between the Pop III IMF and the mass range for PI SNe, the upper IMF limit has only a minor impact on the predicted PI SN rates. If we assume a Pop III IMF with an upper limit of $60\,\Ms$\ the number of Pop III CC SNe increases by almost an order of magnitude.
 \item By making similar assumptions about Pop III SF as the FiBY simulations, we reproduce their PI SN rates.
 \item WDM has a significant impact on the SN rates. By using PI SNe to constrain the onset of primordial SF, it could be possible to discriminate between dark matter models. If Pop III SNe can be detected to redshifts $z\gtrsim 20$, the observational rates will sensitively depend on cosmic structure formation and the critical halo mass for collapse. Even for a small total number of Pop III SN detections ($N\lesssim 10$), high redshift SNe can pose a solid lower limit on the WDM particle mass. Next generation telescopes such as \emph{JWST} and E-ELT should be able to make such observations at high enough redshifts ($z \gtrsim 20$) if Pop III PI SNe are sufficiently abundant. While an observed lack of high redshift SNe could be used to place an upper limit on WDM particle masses, Pop III SF could potentially be delayed e.g. by strong LW feedback, which may produce a similar effect.
\end{itemize}

Applying the proposed diagnostic for structure formation requires that high redshift Pop III PI SNe can be uniquely identified as such. Once detected, a transient source can be identified as a high redshift Pop III PI SN by the slow variation of its light curve (Rydberg et al. in prep.). Could PI SNe at low redshifts \citep[e.g.,][]{gy09,cooke12} be confused with high-$z$ Pop III PI SNe? The progenitors of low-$z$ PI SNe would almost certainly be enriched by metals. \citet{yoon12} find that they could have metallicities as high as 0.3\,$Z_{\odot}$ because stars above this limit would lose so much mass over their lifetimes to line-driven winds that they would never encounter the PI. Those below this metallicity would lose their hydrogen envelopes and explode as bare helium cores, with light curves that are much shorter-lived and dimmer in the NIR than those of Pop III PI SNe, whose progenitors retain their hydrogen envelopes \citep{smidt14a,kz14a,kz14b}. Consequently, PI SNe in the local Universe are easily distinguished from Pop III PI SNe at any redshift.

While we focused on discussing PI SN rates, our models also yield CC SN rates, which have similar profiles in redshift but are four times greater with our
fiducial IMF. Our SN rates can be used to compute expected detection rates for the next generation of ground based and space borne telescopes and to design surveys that maximise the chances of finding the first SNe. The SN rates can quickly be updated for changes in the parameters that govern primordial SF. They can also offer a tool to infer constraints on these parameters from observations of the first SNe.

\section*{Acknowledgements}

We thank the anonymous referee for improving this paper with helpful suggestions and ideas. The authors were supported by the European Research Council under the European Community's Seventh Framework Programme (FP7/2007 - 2013) via the ERC Advanced Grant `STARLIGHT: Formation of the First Stars' under the project number 339177 (MM, RSK, SCOG and DJW) and via the ERC Grant 'BLACK' under the project number 614199 (TH). We thank Jarrett Johnson, John Wise, Michael Norman, Hao Xu, Sadegh Khochfar, Rahul Shetty, Claes-Erik Rydberg, Daniel Ceverino, Anna Schauer and Volker Bromm for insightful comments and discussions.  


\bsp	
\label{lastpage}

\begin{thebibliography}{}
\makeatletter
\relax
\def\mn@urlcharsother{\let\do\@makeother \do\$\do\&\do\#\do\^\do\_\do\%\do\~}
\def\mn@doi{\begingroup\mn@urlcharsother \@ifnextchar [ {\mn@doi@}
  {\mn@doi@[]}}
\def\mn@doi@[#1]#2{\def\@tempa{#1}\ifx\@tempa\@empty \href
  {http://dx.doi.org/#2} {doi:#2}\else \href {http://dx.doi.org/#2} {#1}\fi
  \endgroup}
\def\mn@eprint#1#2{\mn@eprint@#1:#2::\@nil}
\def\mn@eprint@arXiv#1{\href {http://arxiv.org/abs/#1} {{\tt arXiv:#1}}}
\def\mn@eprint@dblp#1{\href {http://dblp.uni-trier.de/rec/bibtex/#1.xml}
  {dblp:#1}}
\def\mn@eprint@#1:#2:#3:#4\@nil{\def\@tempa {#1}\def\@tempb {#2}\def\@tempc
  {#3}\ifx \@tempc \@empty \let \@tempc \@tempb \let \@tempb \@tempa \fi \ifx
  \@tempb \@empty \def\@tempb {arXiv}\fi \@ifundefined
  {mn@eprint@\@tempb}{\@tempb:\@tempc}{\expandafter \expandafter \csname
  mn@eprint@\@tempb\endcsname \expandafter{\@tempc}}}

\bibitem[\protect\citeauthoryear{{Abbott} et~al.,}{{Abbott}
  et~al.}{2016}]{LIGO}
{Abbott} B.~P.,  et~al., 2016, \mn@doi [Physical Review Letters]
  {10.1103/PhysRevLett.116.061102}, \href
  {http://adsabs.harvard.edu/abs/2016PhRvL.116f1102A} {116, 061102}

\bibitem[\protect\citeauthoryear{{Abel}, {Bryan}  \& {Norman}}{{Abel}
  et~al.}{2000}]{abn00}
{Abel} T.,  {Bryan} G.~L.,   {Norman} M.~L.,  2000, \mn@doi [\apj]
  {10.1086/309295}, \href {http://adsabs.harvard.edu/abs/2000ApJ...540...39A}
  {540, 39}

\bibitem[\protect\citeauthoryear{{Abel}, {Bryan}  \& {Norman}}{{Abel}
  et~al.}{2002}]{abn02}
{Abel} T.,  {Bryan} G.~L.,   {Norman} M.~L.,  2002, \mn@doi [Science]
  {10.1126/science.1063991}, \href
  {http://adsabs.harvard.edu/abs/2002Sci...295...93A} {295, 93}

\bibitem[\protect\citeauthoryear{{Abel}, {Wise}  \& {Bryan}}{{Abel}
  et~al.}{2007}]{awb07}
{Abel} T.,  {Wise} J.~H.,   {Bryan} G.~L.,  2007, \mn@doi [\apjl]
  {10.1086/516820}, \href {http://adsabs.harvard.edu/abs/2007ApJ...659L..87A}
  {659, L87}

\bibitem[\protect\citeauthoryear{{Agarwal}, {Khochfar}, {Johnson}, {Neistein},
  {Dalla Vecchia}  \& {Livio}}{{Agarwal} et~al.}{2012}]{agarw12}
{Agarwal} B.,  {Khochfar} S.,  {Johnson} J.~L.,  {Neistein} E.,  {Dalla
  Vecchia} C.,   {Livio} M.,  2012, \mn@doi [\mnras]
  {10.1111/j.1365-2966.2012.21651.x}, \href
  {http://adsabs.harvard.edu/abs/2012MNRAS.425.2854A} {425, 2854}

\bibitem[\protect\citeauthoryear{{Agarwal}, {Dalla Vecchia}, {Johnson},
  {Khochfar}  \& {Paardekooper}}{{Agarwal} et~al.}{2014}]{agarw14}
{Agarwal} B.,  {Dalla Vecchia} C.,  {Johnson} J.~L.,  {Khochfar} S.,
  {Paardekooper} J.-P.,  2014, \mn@doi [\mnras] {10.1093/mnras/stu1112}, \href
  {http://adsabs.harvard.edu/abs/2014MNRAS.443..648A} {443, 648}

\bibitem[\protect\citeauthoryear{{Alvarez}, {Bromm}  \& {Shapiro}}{{Alvarez}
  et~al.}{2006}]{abs06}
{Alvarez} M.~A.,  {Bromm} V.,   {Shapiro} P.~R.,  2006, \mn@doi [\apj]
  {10.1086/499578}, \href {http://adsabs.harvard.edu/abs/2006ApJ...639..621A}
  {639, 621}

\bibitem[\protect\citeauthoryear{{Alvarez}, {Wise}  \& {Abel}}{{Alvarez}
  et~al.}{2009}]{awa09}
{Alvarez} M.~A.,  {Wise} J.~H.,   {Abel} T.,  2009, \mn@doi [\apjl]
  {10.1088/0004-637X/701/2/L133}, \href
  {http://adsabs.harvard.edu/abs/2009ApJ...701L.133A} {701, L133}

\bibitem[\protect\citeauthoryear{{Aoki}, {Tominaga}, {Beers}, {Honda}  \&
  {Lee}}{{Aoki} et~al.}{2014}]{aoki14}
{Aoki} W.,  {Tominaga} N.,  {Beers} T.~C.,  {Honda} S.,   {Lee} Y.~S.,  2014,
  \mn@doi [\sci] {10.1126/science.1252633}, \href
  {http://adsabs.harvard.edu/abs/2014Sci...345..912A} {345, 912}

\bibitem[\protect\citeauthoryear{{Barkana} \& {Loeb}}{{Barkana} \&
  {Loeb}}{2004}]{bl04}
{Barkana} R.,  {Loeb} A.,  2004, \mn@doi [\apj] {10.1086/421079}, \href
  {http://adsabs.harvard.edu/abs/2004ApJ...609..474B} {609, 474}

\bibitem[\protect\citeauthoryear{{Barkat}, {Rakavy}  \& {Sack}}{{Barkat}
  et~al.}{1967}]{brk67}
{Barkat} Z.,  {Rakavy} G.,   {Sack} N.,  1967, \mn@doi [Physical Review
  Letters] {10.1103/PhysRevLett.18.379}, \href
  {http://adsabs.harvard.edu/abs/1967PhRvL..18..379B} {18, 379}

\bibitem[\protect\citeauthoryear{{Beers} \& {Christlieb}}{{Beers} \&
  {Christlieb}}{2005}]{bc05}
{Beers} T.~C.,  {Christlieb} N.,  2005, \mn@doi [\araa]
  {10.1146/annurev.astro.42.053102.134057}, \href
  {http://adsabs.harvard.edu/abs/2005ARA%26A..43..531B} {43, 531}

\bibitem[\protect\citeauthoryear{{Behroozi} \& {Silk}}{{Behroozi} \&
  {Silk}}{2015}]{BehrooziSilk2015}
{Behroozi} P.~S.,  {Silk} J.,  2015, \mn@doi [\apj]
  {10.1088/0004-637X/799/1/32}, \href
  {http://adsabs.harvard.edu/abs/2015ApJ...799...32B} {799, 32}

\bibitem[\protect\citeauthoryear{{Bond}, {Arnett}  \& {Carr}}{{Bond}
  et~al.}{1984}]{bet84}
{Bond} J.~R.,  {Arnett} W.~D.,   {Carr} B.~J.,  1984, \mn@doi [\apj]
  {10.1086/162057}, \href {http://adsabs.harvard.edu/abs/1984ApJ...280..825B}
  {280, 825}

\bibitem[\protect\citeauthoryear{{Bond}, {Cole}, {Efstathiou}  \&
  {Kaiser}}{{Bond} et~al.}{1991}]{Bond1991}
{Bond} J.~R.,  {Cole} S.,  {Efstathiou} G.,   {Kaiser} N.,  1991, \mn@doi
  [\apj] {10.1086/170520}, \href
  {http://adsabs.harvard.edu/abs/1991ApJ...379..440B} {379, 440}

\bibitem[\protect\citeauthoryear{{Bromm}}{{Bromm}}{2013}]{BrommReview}
{Bromm} V.,  2013, \mn@doi [Reports on Progress in Physics]
  {10.1088/0034-4885/76/11/112901}, \href
  {http://adsabs.harvard.edu/abs/2013RPPh...76k2901B} {76, 112901}

\bibitem[\protect\citeauthoryear{{Bromm} \& {Yoshida}}{{Bromm} \&
  {Yoshida}}{2011}]{fg11}
{Bromm} V.,  {Yoshida} N.,  2011, \mn@doi [\araa]
  {10.1146/annurev-astro-081710-102608}, \href
  {http://adsabs.harvard.edu/abs/2011ARA%26A..49..373B} {49, 373}

\bibitem[\protect\citeauthoryear{{Bromm}, {Coppi}  \& {Larson}}{{Bromm}
  et~al.}{1999}]{bcl99}
{Bromm} V.,  {Coppi} P.~S.,   {Larson} R.~B.,  1999, \mn@doi [\apjl]
  {10.1086/312385}, \href {http://adsabs.harvard.edu/abs/1999ApJ...527L...5B}
  {527, L5}

\bibitem[\protect\citeauthoryear{{Bromm}, {Coppi}  \& {Larson}}{{Bromm}
  et~al.}{2002}]{bcl02}
{Bromm} V.,  {Coppi} P.~S.,   {Larson} R.~B.,  2002, \mn@doi [\apj]
  {10.1086/323947}, \href {http://adsabs.harvard.edu/abs/2002ApJ...564...23B}
  {564, 23}

\bibitem[\protect\citeauthoryear{{Bromm}, {Yoshida}  \& {Hernquist}}{{Bromm}
  et~al.}{2003}]{byh03}
{Bromm} V.,  {Yoshida} N.,   {Hernquist} L.,  2003, \mn@doi [\apjl]
  {10.1086/379359}, \href {http://adsabs.harvard.edu/abs/2003ApJ...596L.135B}
  {596, L135}

\bibitem[\protect\citeauthoryear{{Bromm}, {Yoshida}, {Hernquist}  \&
  {McKee}}{{Bromm} et~al.}{2009}]{fsg09}
{Bromm} V.,  {Yoshida} N.,  {Hernquist} L.,   {McKee} C.~F.,  2009, \mn@doi
  [\nat] {10.1038/nature07990}, \href
  {http://adsabs.harvard.edu/abs/2009Natur.459...49B} {459, 49}

\bibitem[\protect\citeauthoryear{{Cen} \& {Riquelme}}{{Cen} \&
  {Riquelme}}{2008}]{Cen2008}
{Cen} R.,  {Riquelme} M.~A.,  2008, \mn@doi [\apj] {10.1086/524724}, \href
  {http://adsabs.harvard.edu/abs/2008ApJ...674..644C} {674, 644}

\bibitem[\protect\citeauthoryear{{Chabrier}}{{Chabrier}}{2003}]{ChabrierIMF}
{Chabrier} G.,  2003, \mn@doi [\pasp] {10.1086/376392}, \href
  {http://ads.ari.uni-heidelberg.de/abs/2003PASP..115..763C} {115, 763}

\bibitem[\protect\citeauthoryear{{Chatzopoulos} \& {Wheeler}}{{Chatzopoulos} \&
  {Wheeler}}{2012}]{cw12}
{Chatzopoulos} E.,  {Wheeler} J.~C.,  2012, \mn@doi [\apj]
  {10.1088/0004-637X/748/1/42}, \href
  {http://adsabs.harvard.edu/abs/2012ApJ...748...42C} {748, 42}

\bibitem[\protect\citeauthoryear{{Chatzopoulos}, {Wheeler}  \&
  {Couch}}{{Chatzopoulos} et~al.}{2013}]{cwc13}
{Chatzopoulos} E.,  {Wheeler} J.~C.,   {Couch} S.~M.,  2013, \mn@doi [\apj]
  {10.1088/0004-637X/776/2/129}, \href
  {http://adsabs.harvard.edu/abs/2013ApJ...776..129C} {776, 129}

\bibitem[\protect\citeauthoryear{{Chatzopoulos}, {van Rossum}, {Craig},
  {Whalen}, {Smidt}  \& {Wiggins}}{{Chatzopoulos} et~al.}{2015}]{CW15}
{Chatzopoulos} E.,  {van Rossum} D.~R.,  {Craig} W.~J.,  {Whalen} D.~J.,
  {Smidt} J.,   {Wiggins} B.,  2015, \mn@doi [\apj]
  {10.1088/0004-637X/799/1/18}, \href
  {http://ads.ari.uni-heidelberg.de/abs/2015ApJ...799...18C} {799, 18}

\bibitem[\protect\citeauthoryear{{Chen}, {Woosley}, {Heger}, {Almgren}  \&
  {Whalen}}{{Chen} et~al.}{2014a}]{chen14a}
{Chen} K.-J.,  {Woosley} S.,  {Heger} A.,  {Almgren} A.,   {Whalen} D.~J.,
  2014a, \mn@doi [\apj] {10.1088/0004-637X/792/1/28}, \href
  {http://adsabs.harvard.edu/abs/2014ApJ...792...28C} {792, 28}

\bibitem[\protect\citeauthoryear{{Chen}, {Heger}, {Woosley}, {Almgren}  \&
  {Whalen}}{{Chen} et~al.}{2014b}]{chen14c}
{Chen} K.-J.,  {Heger} A.,  {Woosley} S.,  {Almgren} A.,   {Whalen} D.~J.,
  2014b, \mn@doi [\apj] {10.1088/0004-637X/792/1/44}, \href
  {http://adsabs.harvard.edu/abs/2014ApJ...792...44C} {792, 44}

\bibitem[\protect\citeauthoryear{{Chen}, {Heger}, {Whalen}, {Moriya}, {Bromm},
  {Woosley}  \& {Almgren}}{{Chen} et~al.}{2016}]{Chen16}
{Chen} K.-J.,  {Heger} A.,  {Whalen} D.~J.,  {Moriya} T.~J.,  {Bromm} V.,
  {Woosley} S.,   {Almgren} A.,  2016, preprint, \href
  {http://adsabs.harvard.edu/abs/2016arXiv160106896C} {} (\mn@eprint {arXiv}
  {1601.06896})

\bibitem[\protect\citeauthoryear{{Clark}, {Glover}, {Smith}, {Greif}, {Klessen}
   \& {Bromm}}{{Clark} et~al.}{2011}]{clark11}
{Clark} P.~C.,  {Glover} S.~C.~O.,  {Smith} R.~J.,  {Greif} T.~H.,  {Klessen}
  R.~S.,   {Bromm} V.,  2011, \mn@doi [Science] {10.1126/science.1198027},
  \href {http://adsabs.harvard.edu/abs/2011Sci...331.1040C} {331, 1040}

\bibitem[\protect\citeauthoryear{{Cole}, {Lacey}, {Baugh}  \& {Frenk}}{{Cole}
  et~al.}{2000}]{galform}
{Cole} S.,  {Lacey} C.~G.,  {Baugh} C.~M.,   {Frenk} C.~S.,  2000, \mn@doi
  [\mnras] {10.1046/j.1365-8711.2000.03879.x}, \href
  {http://adsabs.harvard.edu/abs/2000MNRAS.319..168C} {319, 168}

\bibitem[\protect\citeauthoryear{{Cooke} et~al.,}{{Cooke}
  et~al.}{2012}]{cooke12}
{Cooke} J.,  et~al., 2012, \mn@doi [\nat] {10.1038/nature11521}, \href
  {http://adsabs.harvard.edu/abs/2012Natur.491..228C} {491, 228}

\bibitem[\protect\citeauthoryear{{Crowther}, {Schnurr}, {Hirschi}, {Yusof},
  {Parker}, {Goodwin}  \& {Kassim}}{{Crowther} et~al.}{2010}]{R136}
{Crowther} P.~A.,  {Schnurr} O.,  {Hirschi} R.,  {Yusof} N.,  {Parker} R.~J.,
  {Goodwin} S.~P.,   {Kassim} H.~A.,  2010, \mn@doi [\mnras]
  {10.1111/j.1365-2966.2010.17167.x}, \href
  {http://adsabs.harvard.edu/abs/2010MNRAS.408..731C} {408, 731}

\bibitem[\protect\citeauthoryear{{D'Onghia} \& {Lake}}{{D'Onghia} \&
  {Lake}}{2004}]{DOnghia2004}
{D'Onghia} E.,  {Lake} G.,  2004, \mn@doi [\apj] {10.1086/422794}, \href
  {http://adsabs.harvard.edu/abs/2004ApJ...612..628D} {612, 628}

\bibitem[\protect\citeauthoryear{{Dopcke}, {Glover}, {Clark}  \&
  {Klessen}}{{Dopcke} et~al.}{2013}]{Dopcke13}
{Dopcke} G.,  {Glover} S.~C.~O.,  {Clark} P.~C.,   {Klessen} R.~S.,  2013,
  \mn@doi [\apj] {10.1088/0004-637X/766/2/103}, \href
  {http://ads.ari.uni-heidelberg.de/abs/2013ApJ...766..103D} {766, 103}

\bibitem[\protect\citeauthoryear{{Dvorkin}, {Vangioni}, {Silk}, {Uzan}  \&
  {Olive}}{{Dvorkin} et~al.}{2016}]{Dvorkin16}
{Dvorkin} I.,  {Vangioni} E.,  {Silk} J.,  {Uzan} J.-P.,   {Olive} K.~A.,
  2016, preprint, \href {http://adsabs.harvard.edu/abs/2016arXiv160404288D} {}
  (\mn@eprint {arXiv} {1604.04288})

\bibitem[\protect\citeauthoryear{{Frebel} et~al.,}{{Frebel}
  et~al.}{2005}]{fet05}
{Frebel} A.,  et~al., 2005, \mn@doi [\nat] {10.1038/nature03455}, \href
  {http://adsabs.harvard.edu/abs/2005Natur.434..871F} {434, 871}

\bibitem[\protect\citeauthoryear{{Frebel}, {Johnson}  \& {Bromm}}{{Frebel}
  et~al.}{2008}]{Frebel08}
{Frebel} A.,  {Johnson} J.~L.,   {Bromm} V.,  2008, in {Hunt} L.~K.,  {Madden}
  S.~C.,   {Schneider} R.,  eds,  IAU Symposium Vol. 255, Low-Metallicity Star
  Formation: From the First Stars to Dwarf Galaxies. pp 336--340,
  \mn@doi{10.1017/S1743921308025039}

\bibitem[\protect\citeauthoryear{{Gal-Yam} et~al.,}{{Gal-Yam}
  et~al.}{2009}]{gy09}
{Gal-Yam} A.,  et~al., 2009, \mn@doi [\nat] {10.1038/nature08579}, \href
  {http://adsabs.harvard.edu/abs/2009Natur.462..624G} {462, 624}

\bibitem[\protect\citeauthoryear{{Gardner} et~al.,}{{Gardner}
  et~al.}{2006}]{jwst06}
{Gardner} J.~P.,  et~al., 2006, \mn@doi [\ssr] {10.1007/s11214-006-8315-7},
  \href {http://adsabs.harvard.edu/abs/2006SSRv..123..485G} {123, 485}

\bibitem[\protect\citeauthoryear{{Gilmozzi} \& {Spyromilio}}{{Gilmozzi} \&
  {Spyromilio}}{2007}]{EELT07}
{Gilmozzi} R.,  {Spyromilio} J.,  2007, The Messenger, \href
  {http://adsabs.harvard.edu/abs/2007Msngr.127...11G} {127}

\bibitem[\protect\citeauthoryear{{Glover}}{{Glover}}{2013}]{GloverReview}
{Glover} S.,  2013, in {Wiklind} T.,  {Mobasher} B.,   {Bromm} V.,  eds,
  Astrophysics and Space Science Library Vol. 396, Astrophysics and Space
  Science Library. p.~103 (\mn@eprint {arXiv} {1209.2509}),
  \mn@doi{10.1007/978-3-642-32362-1_3}

\bibitem[\protect\citeauthoryear{{Gou}, {M{\'e}sz{\'a}ros}, {Abel}  \&
  {Zhang}}{{Gou} et~al.}{2004}]{gou04}
{Gou} L.~J.,  {M{\'e}sz{\'a}ros} P.,  {Abel} T.,   {Zhang} B.,  2004, \mn@doi
  [\apj] {10.1086/382061}, \href
  {http://adsabs.harvard.edu/abs/2004ApJ...604..508G} {604, 508}

\bibitem[\protect\citeauthoryear{{Greif}, {White}, {Klessen}  \&
  {Springel}}{{Greif} et~al.}{2011a}]{Greif2011a}
{Greif} T.~H.,  {White} S.~D.~M.,  {Klessen} R.~S.,   {Springel} V.,  2011a,
  \mn@doi [\apj] {10.1088/0004-637X/736/2/147}, \href
  {http://adsabs.harvard.edu/abs/2011ApJ...736..147G} {736, 147}

\bibitem[\protect\citeauthoryear{{Greif}, {Springel}, {White}, {Glover},
  {Clark}, {Smith}, {Klessen}  \& {Bromm}}{{Greif} et~al.}{2011b}]{Greif11b}
{Greif} T.~H.,  {Springel} V.,  {White} S.~D.~M.,  {Glover} S.~C.~O.,  {Clark}
  P.~C.,  {Smith} R.~J.,  {Klessen} R.~S.,   {Bromm} V.,  2011b, \mn@doi [\apj]
  {10.1088/0004-637X/737/2/75}, \href
  {http://adsabs.harvard.edu/abs/2011ApJ...737...75G} {737, 75}

\bibitem[\protect\citeauthoryear{{Greif}, {Bromm}, {Clark}, {Glover}, {Smith},
  {Klessen}, {Yoshida}  \& {Springel}}{{Greif} et~al.}{2012}]{get12}
{Greif} T.~H.,  {Bromm} V.,  {Clark} P.~C.,  {Glover} S.~C.~O.,  {Smith} R.~J.,
   {Klessen} R.~S.,  {Yoshida} N.,   {Springel} V.,  2012, \mn@doi [\mnras]
  {10.1111/j.1365-2966.2012.21212.x}, \href
  {http://adsabs.harvard.edu/abs/2012MNRAS.424..399G} {424, 399}

\bibitem[\protect\citeauthoryear{{Griffen}, {Ji}, {Dooley}, {G{\'o}mez},
  {Vogelsberger}, {O'Shea}  \& {Frebel}}{{Griffen} et~al.}{2016}]{Caterpillar}
{Griffen} B.~F.,  {Ji} A.~P.,  {Dooley} G.~A.,  {G{\'o}mez} F.~A.,
  {Vogelsberger} M.,  {O'Shea} B.~W.,   {Frebel} A.,  2016, \mn@doi [\apj]
  {10.3847/0004-637X/818/1/10}, \href
  {http://adsabs.harvard.edu/abs/2016ApJ...818...10G} {818, 10}

\bibitem[\protect\citeauthoryear{{Hartwig} et~al.,}{{Hartwig}
  et~al.}{2015a}]{Hartwig2016a}
{Hartwig} T.,  et~al., 2015a, preprint (\mn@eprint {arXiv} {1512.01111})

\bibitem[\protect\citeauthoryear{{Hartwig}, {Bromm}, {Klessen}  \&
  {Glover}}{{Hartwig} et~al.}{2015b}]{Hartwig15b}
{Hartwig} T.,  {Bromm} V.,  {Klessen} R.~S.,   {Glover} S.~C.~O.,  2015b,
  \mn@doi [\mnras] {10.1093/mnras/stu2740}, \href
  {http://adsabs.harvard.edu/abs/2015MNRAS.447.3892H} {447, 3892}

\bibitem[\protect\citeauthoryear{{Hartwig}, {Clark}, {Glover}, {Klessen}  \&
  {Sasaki}}{{Hartwig} et~al.}{2015c}]{Hartwig15a}
{Hartwig} T.,  {Clark} P.~C.,  {Glover} S.~C.~O.,  {Klessen} R.~S.,   {Sasaki}
  M.,  2015c, \mn@doi [\apj] {10.1088/0004-637X/799/2/114}, \href
  {http://adsabs.harvard.edu/abs/2015ApJ...799..114H} {799, 114}

\bibitem[\protect\citeauthoryear{{Hartwig}, {Volonteri}, {Bromm}, {Klessen},
  {Barausse}, {Magg}  \& {Stacy}}{{Hartwig} et~al.}{2016}]{Hartwig2016b}
{Hartwig} T.,  {Volonteri} M.,  {Bromm} V.,  {Klessen} R.~S.,  {Barausse} E.,
  {Magg} M.,   {Stacy} A.,  2016, \mn@doi [\mnras] {10.1093/mnrasl/slw074},
  \href {http://adsabs.harvard.edu/abs/2016MNRAS.460L..74H} {460, L74}

\bibitem[\protect\citeauthoryear{{Heger} \& {Woosley}}{{Heger} \&
  {Woosley}}{2002}]{HegerWoosley2002}
{Heger} A.,  {Woosley} S.~E.,  2002, \mn@doi [\apj] {10.1086/338487}, \href
  {http://adsabs.harvard.edu/abs/2002ApJ...567..532H} {567, 532}

\bibitem[\protect\citeauthoryear{{Hirano}, {Hosokawa}, {Yoshida}, {Umeda},
  {Omukai}, {Chiaki}  \& {Yorke}}{{Hirano} et~al.}{2014}]{hir13}
{Hirano} S.,  {Hosokawa} T.,  {Yoshida} N.,  {Umeda} H.,  {Omukai} K.,
  {Chiaki} G.,   {Yorke} H.~W.,  2014, \mn@doi [\apj]
  {10.1088/0004-637X/781/2/60}, \href
  {http://adsabs.harvard.edu/abs/2014ApJ...781...60H} {781, 60}

\bibitem[\protect\citeauthoryear{{Hosokawa}, {Omukai}, {Yoshida}  \&
  {Yorke}}{{Hosokawa} et~al.}{2011}]{hos11}
{Hosokawa} T.,  {Omukai} K.,  {Yoshida} N.,   {Yorke} H.~W.,  2011, \mn@doi
  [Science] {10.1126/science.1207433}, \href
  {http://adsabs.harvard.edu/abs/2011Sci...334.1250H} {334, 1250}

\bibitem[\protect\citeauthoryear{{Hummel}, {Pawlik}, {Milosavljevi{\'c}}  \&
  {Bromm}}{{Hummel} et~al.}{2012}]{hum12}
{Hummel} J.~A.,  {Pawlik} A.~H.,  {Milosavljevi{\'c}} M.,   {Bromm} V.,  2012,
  \mn@doi [\apj] {10.1088/0004-637X/755/1/72}, \href
  {http://adsabs.harvard.edu/abs/2012ApJ...755...72H} {755, 72}

\bibitem[\protect\citeauthoryear{{Inayoshi}, {Kashiyama}, {Visbal}  \&
  {Haiman}}{{Inayoshi} et~al.}{2016}]{Inayoshi16}
{Inayoshi} K.,  {Kashiyama} K.,  {Visbal} E.,   {Haiman} Z.,  2016, preprint,
  \href {http://adsabs.harvard.edu/abs/2016arXiv160306921I} {} (\mn@eprint
  {arXiv} {1603.06921})

\bibitem[\protect\citeauthoryear{{Ishiyama}, {Sudo}, {Yokoi}, {Hasegawa},
  {Tominaga}  \& {Susa}}{{Ishiyama} et~al.}{2016}]{Ishiyama16}
{Ishiyama} T.,  {Sudo} K.,  {Yokoi} S.,  {Hasegawa} K.,  {Tominaga} N.,
  {Susa} H.,  2016, preprint, \href
  {http://adsabs.harvard.edu/abs/2016arXiv160200465I} {} (\mn@eprint {arXiv}
  {1602.00465})

\bibitem[\protect\citeauthoryear{{Jiang} \& {van den Bosch}}{{Jiang} \& {van
  den Bosch}}{2014}]{Jiang2014}
{Jiang} F.,  {van den Bosch} F.~C.,  2014, \mn@doi [\mnras]
  {10.1093/mnras/stu280}, \href
  {http://adsabs.harvard.edu/abs/2014MNRAS.440..193J} {440, 193}

\bibitem[\protect\citeauthoryear{{Joggerst} \& {Whalen}}{{Joggerst} \&
  {Whalen}}{2011}]{jw11}
{Joggerst} C.~C.,  {Whalen} D.~J.,  2011, \mn@doi [\apj]
  {10.1088/0004-637X/728/2/129}, \href
  {http://adsabs.harvard.edu/abs/2011ApJ...728..129J} {728, 129}

\bibitem[\protect\citeauthoryear{{Joggerst}, {Almgren}, {Bell}, {Heger},
  {Whalen}  \& {Woosley}}{{Joggerst} et~al.}{2010}]{jet09b}
{Joggerst} C.~C.,  {Almgren} A.,  {Bell} J.,  {Heger} A.,  {Whalen} D.,
  {Woosley} S.~E.,  2010, \mn@doi [\apj] {10.1088/0004-637X/709/1/11}, \href
  {http://adsabs.harvard.edu/abs/2010ApJ...709...11J} {709, 11}

\bibitem[\protect\citeauthoryear{{Johnson}, {Whalen}, {Fryer}  \&
  {Li}}{{Johnson} et~al.}{2012}]{jlj12a}
{Johnson} J.~L.,  {Whalen} D.~J.,  {Fryer} C.~L.,   {Li} H.,  2012, \mn@doi
  [\apj] {10.1088/0004-637X/750/1/66}, \href
  {http://adsabs.harvard.edu/abs/2012ApJ...750...66J} {750, 66}

\bibitem[\protect\citeauthoryear{{Johnson}, {Dalla}  \& {Khochfar}}{{Johnson}
  et~al.}{2013a}]{FiBY1}
{Johnson} J.~L.,  {Dalla} V.~C.,   {Khochfar} S.,  2013a, \mn@doi [\mnras]
  {10.1093/mnras/sts011}, \href
  {http://adsabs.harvard.edu/abs/2013MNRAS.428.1857J} {428, 1857}

\bibitem[\protect\citeauthoryear{{Johnson}, {Whalen}, {Li}  \&
  {Holz}}{{Johnson} et~al.}{2013b}]{jet13}
{Johnson} J.~L.,  {Whalen} D.~J.,  {Li} H.,   {Holz} D.~E.,  2013b, \mn@doi
  [\apj] {10.1088/0004-637X/771/2/116}, \href
  {http://adsabs.harvard.edu/abs/2013ApJ...771..116J} {771, 116}

\bibitem[\protect\citeauthoryear{{Johnson}, {Whalen}, {Agarwal}, {Paardekooper}
   \& {Khochfar}}{{Johnson} et~al.}{2014}]{jet14}
{Johnson} J.~L.,  {Whalen} D.~J.,  {Agarwal} B.,  {Paardekooper} J.-P.,
  {Khochfar} S.,  2014, \mn@doi [\mnras] {10.1093/mnras/stu1676}, \href
  {http://adsabs.harvard.edu/abs/2014MNRAS.445..686J} {445, 686}

\bibitem[\protect\citeauthoryear{{Karlsson}, {Bromm}  \&
  {Bland-Hawthorn}}{{Karlsson} et~al.}{2013}]{KarlssonReview}
{Karlsson} T.,  {Bromm} V.,   {Bland-Hawthorn} J.,  2013, \mn@doi [Reviews of
  Modern Physics] {10.1103/RevModPhys.85.809}, \href
  {http://adsabs.harvard.edu/abs/2013RvMP...85..809K} {85, 809}

\bibitem[\protect\citeauthoryear{{Kasen}, {Woosley}  \& {Heger}}{{Kasen}
  et~al.}{2011}]{kasen11}
{Kasen} D.,  {Woosley} S.~E.,   {Heger} A.,  2011, \mn@doi [\apj]
  {10.1088/0004-637X/734/2/102}, \href
  {http://adsabs.harvard.edu/abs/2011ApJ...734..102K} {734, 102}

\bibitem[\protect\citeauthoryear{{Kitayama}, {Yoshida}, {Susa}  \&
  {Umemura}}{{Kitayama} et~al.}{2004}]{ket04}
{Kitayama} T.,  {Yoshida} N.,  {Susa} H.,   {Umemura} M.,  2004, \mn@doi [\apj]
  {10.1086/423313}, \href {http://adsabs.harvard.edu/abs/2004ApJ...613..631K}
  {613, 631}

\bibitem[\protect\citeauthoryear{{Komatsu} et~al.,}{{Komatsu}
  et~al.}{2011}]{wmap7year}
{Komatsu} E.,  et~al., 2011, \mn@doi [\apjs] {10.1088/0067-0049/192/2/18},
  \href {http://adsabs.harvard.edu/abs/2011ApJS..192...18K} {192, 18}

\bibitem[\protect\citeauthoryear{{Kozyreva}, {Blinnikov}, {Langer}  \&
  {Yoon}}{{Kozyreva} et~al.}{2014a}]{kz14a}
{Kozyreva} A.,  {Blinnikov} S.,  {Langer} N.,   {Yoon} S.-C.,  2014a, \mn@doi
  [\aap] {10.1051/0004-6361/201423447}, \href
  {http://adsabs.harvard.edu/abs/2014A%26A...565A..70K} {565, A70}

\bibitem[\protect\citeauthoryear{{Kozyreva}, {Yoon}  \& {Langer}}{{Kozyreva}
  et~al.}{2014b}]{kz14b}
{Kozyreva} A.,  {Yoon} S.-C.,   {Langer} N.,  2014b, \mn@doi [\aap]
  {10.1051/0004-6361/201423641}, \href
  {http://adsabs.harvard.edu/abs/2014A%26A...566A.146K} {566, A146}

\bibitem[\protect\citeauthoryear{{Kroupa}}{{Kroupa}}{2001}]{KroupaIMF}
{Kroupa} P.,  2001, \mn@doi [\mnras] {10.1046/j.1365-8711.2001.04022.x}, \href
  {http://ads.ari.uni-heidelberg.de/abs/2001MNRAS.322..231K} {322, 231}

\bibitem[\protect\citeauthoryear{{Kroupa}}{{Kroupa}}{2012}]{Kroupa12}
{Kroupa} P.,  2012, \mn@doi [\pasa] {10.1071/AS12005}, \href
  {http://adsabs.harvard.edu/abs/2012PASA...29..395K} {29, 395}

\bibitem[\protect\citeauthoryear{{Lacey} \& {Cole}}{{Lacey} \&
  {Cole}}{1993}]{LaceyCole1993}
{Lacey} C.,  {Cole} S.,  1993, \mnras, \href
  {http://adsabs.harvard.edu/abs/1993MNRAS.262..627L} {262, 627}

\bibitem[\protect\citeauthoryear{{Latif}, {Schleicher}, {Schmidt}  \&
  {Niemeyer}}{{Latif} et~al.}{2013a}]{latif13a}
{Latif} M.~A.,  {Schleicher} D.~R.~G.,  {Schmidt} W.,   {Niemeyer} J.,  2013a,
  \mn@doi [\mnras] {10.1093/mnras/sts659}, \href
  {http://adsabs.harvard.edu/abs/2013MNRAS.430..588L} {430, 588}

\bibitem[\protect\citeauthoryear{{Latif}, {Schleicher}, {Schmidt}  \&
  {Niemeyer}}{{Latif} et~al.}{2013b}]{latif13c}
{Latif} M.~A.,  {Schleicher} D.~R.~G.,  {Schmidt} W.,   {Niemeyer} J.,  2013b,
  \mn@doi [\mnras] {10.1093/mnras/stt834}, \href
  {http://adsabs.harvard.edu/abs/2013MNRAS.433.1607L} {433, 1607}

\bibitem[\protect\citeauthoryear{{Laureijs} et~al.,}{{Laureijs}
  et~al.}{2011}]{euclid}
{Laureijs} R.,  et~al., 2011, preprint (\mn@eprint {arXiv} {1110.3193})

\bibitem[\protect\citeauthoryear{{Machacek}, {Bryan}  \& {Abel}}{{Machacek}
  et~al.}{2001}]{Machacek2001}
{Machacek} M.~E.,  {Bryan} G.~L.,   {Abel} T.,  2001, \mn@doi [\apj]
  {10.1086/319014}, \href {http://adsabs.harvard.edu/abs/2001ApJ...548..509M}
  {548, 509}

\bibitem[\protect\citeauthoryear{{Mackey}, {Bromm}  \& {Hernquist}}{{Mackey}
  et~al.}{2003}]{mbh03}
{Mackey} J.,  {Bromm} V.,   {Hernquist} L.,  2003, \mn@doi [\apj]
  {10.1086/367613}, \href {http://adsabs.harvard.edu/abs/2003ApJ...586....1M}
  {586, 1}

\bibitem[\protect\citeauthoryear{{Madau} \& {Haardt}}{{Madau} \&
  {Haardt}}{2015}]{Madau15}
{Madau} P.,  {Haardt} F.,  2015, \mn@doi [\apjl] {10.1088/2041-8205/813/1/L8},
  \href {http://adsabs.harvard.edu/abs/2015ApJ...813L...8M} {813, L8}

\bibitem[\protect\citeauthoryear{{Maio} \& {Viel}}{{Maio} \&
  {Viel}}{2015}]{Maio2015}
{Maio} U.,  {Viel} M.,  2015, \mn@doi [\mnras] {10.1093/mnras/stu2304}, \href
  {http://adsabs.harvard.edu/abs/2015MNRAS.446.2760M} {446, 2760}

\bibitem[\protect\citeauthoryear{{Maio}, {Koopmans}  \& {Ciardi}}{{Maio}
  et~al.}{2011}]{Maio2011}
{Maio} U.,  {Koopmans} L.~V.~E.,   {Ciardi} B.,  2011, \mn@doi [\mnras]
  {10.1111/j.1745-3933.2010.01001.x}, \href
  {http://adsabs.harvard.edu/abs/2011MNRAS.412L..40M} {412, L40}

\bibitem[\protect\citeauthoryear{{Mesler}, {Whalen}, {Smidt}, {Fryer},
  {Lloyd-Ronning}  \& {Pihlstr{\"o}m}}{{Mesler} et~al.}{2014}]{mes13a}
{Mesler} R.~A.,  {Whalen} D.~J.,  {Smidt} J.,  {Fryer} C.~L.,  {Lloyd-Ronning}
  N.~M.,   {Pihlstr{\"o}m} Y.~M.,  2014, \mn@doi [\apj]
  {10.1088/0004-637X/787/1/91}, \href
  {http://adsabs.harvard.edu/abs/2014ApJ...787...91M} {787, 91}

\bibitem[\protect\citeauthoryear{{M{\'e}sz{\'a}ros} \&
  {Rees}}{{M{\'e}sz{\'a}ros} \& {Rees}}{2010}]{mesz10}
{M{\'e}sz{\'a}ros} P.,  {Rees} M.~J.,  2010, \mn@doi [\apj]
  {10.1088/0004-637X/715/2/967}, \href
  {http://adsabs.harvard.edu/abs/2010ApJ...715..967M} {715, 967}

\bibitem[\protect\citeauthoryear{{Milosavljevi{\'c}}, {Couch}  \&
  {Bromm}}{{Milosavljevi{\'c}} et~al.}{2009a}]{milos09a}
{Milosavljevi{\'c}} M.,  {Couch} S.~M.,   {Bromm} V.,  2009a, \mn@doi [\apjl]
  {10.1088/0004-637X/696/2/L146}, \href
  {http://adsabs.harvard.edu/abs/2009ApJ...696L.146M} {696, L146}

\bibitem[\protect\citeauthoryear{{Milosavljevi{\'c}}, {Bromm}, {Couch}  \&
  {Oh}}{{Milosavljevi{\'c}} et~al.}{2009b}]{milos09b}
{Milosavljevi{\'c}} M.,  {Bromm} V.,  {Couch} S.~M.,   {Oh} S.~P.,  2009b,
  \mn@doi [\apj] {10.1088/0004-637X/698/1/766}, \href
  {http://adsabs.harvard.edu/abs/2009ApJ...698..766M} {698, 766}

\bibitem[\protect\citeauthoryear{{Moriya}, {Blinnikov}, {Tominaga}, {Yoshida},
  {Tanaka}, {Maeda}  \& {Nomoto}}{{Moriya} et~al.}{2013}]{moriya12}
{Moriya} T.~J.,  {Blinnikov} S.~I.,  {Tominaga} N.,  {Yoshida} N.,  {Tanaka}
  M.,  {Maeda} K.,   {Nomoto} K.,  2013, \mn@doi [\mnras]
  {10.1093/mnras/sts075}, \href
  {http://adsabs.harvard.edu/abs/2013MNRAS.428.1020M} {428, 1020}

\bibitem[\protect\citeauthoryear{{Murray}, {Power}  \& {Robotham}}{{Murray}
  et~al.}{2013}]{HMFcalc}
{Murray} S.~G.,  {Power} C.,   {Robotham} A.~S.~G.,  2013, \mn@doi [Astronomy
  and Computing] {10.1016/j.ascom.2013.11.001}, \href
  {http://adsabs.harvard.edu/abs/2013A%26C.....3...23M} {3, 23}

\bibitem[\protect\citeauthoryear{{Nakamura} \& {Umemura}}{{Nakamura} \&
  {Umemura}}{2001}]{nu01}
{Nakamura} F.,  {Umemura} M.,  2001, \mn@doi [\apj] {10.1086/318663}, \href
  {http://adsabs.harvard.edu/abs/2001ApJ...548...19N} {548, 19}

\bibitem[\protect\citeauthoryear{{O'Shea} \& {Norman}}{{O'Shea} \&
  {Norman}}{2007}]{on07}
{O'Shea} B.~W.,  {Norman} M.~L.,  2007, \mn@doi [\apj] {10.1086/509250}, \href
  {http://adsabs.harvard.edu/abs/2007ApJ...654...66O} {654, 66}

\bibitem[\protect\citeauthoryear{{O'Shea}, {Wise}, {Xu}  \& {Norman}}{{O'Shea}
  et~al.}{2015}]{ren15}
{O'Shea} B.~W.,  {Wise} J.~H.,  {Xu} H.,   {Norman} M.~L.,  2015, \mn@doi
  [\apjl] {10.1088/2041-8205/807/1/L12}, \href
  {http://adsabs.harvard.edu/abs/2015ApJ...807L..12O} {807, L12}

\bibitem[\protect\citeauthoryear{{Paardekooper}, {Khochfar}  \& {Dalla
  Vecchia}}{{Paardekooper} et~al.}{2013}]{Paardekooper2013}
{Paardekooper} J.-P.,  {Khochfar} S.,   {Dalla Vecchia} C.,  2013, \mn@doi
  [\mnras] {10.1093/mnrasl/sls032}, \href
  {http://adsabs.harvard.edu/abs/2013MNRAS.429L..94P} {429, L94}

\bibitem[\protect\citeauthoryear{{Pan}, {Kasen}  \& {Loeb}}{{Pan}
  et~al.}{2012}]{pan12a}
{Pan} T.,  {Kasen} D.,   {Loeb} A.,  2012, \mn@doi [\mnras]
  {10.1111/j.1365-2966.2012.20837.x}, \href
  {http://adsabs.harvard.edu/abs/2012MNRAS.422.2701P} {422, 2701}

\bibitem[\protect\citeauthoryear{{Park} \& {Ricotti}}{{Park} \&
  {Ricotti}}{2011}]{pm11}
{Park} K.,  {Ricotti} M.,  2011, \mn@doi [\apj] {10.1088/0004-637X/739/1/2},
  \href {http://adsabs.harvard.edu/abs/2011ApJ...739....2P} {739, 2}

\bibitem[\protect\citeauthoryear{{Park} \& {Ricotti}}{{Park} \&
  {Ricotti}}{2012}]{pm12}
{Park} K.,  {Ricotti} M.,  2012, \mn@doi [\apj] {10.1088/0004-637X/747/1/9},
  \href {http://adsabs.harvard.edu/abs/2012ApJ...747....9P} {747, 9}

\bibitem[\protect\citeauthoryear{{Park} \& {Ricotti}}{{Park} \&
  {Ricotti}}{2013}]{pm13}
{Park} K.,  {Ricotti} M.,  2013, \mn@doi [\apj] {10.1088/0004-637X/767/2/163},
  \href {http://adsabs.harvard.edu/abs/2013ApJ...767..163P} {767, 163}

\bibitem[\protect\citeauthoryear{{Parkinson}, {Cole}  \& {Helly}}{{Parkinson}
  et~al.}{2008}]{Parkinson2008}
{Parkinson} H.,  {Cole} S.,   {Helly} J.,  2008, \mn@doi [\mnras]
  {10.1111/j.1365-2966.2007.12517.x}, \href
  {http://adsabs.harvard.edu/abs/2008MNRAS.383..557P} {383, 557}

\bibitem[\protect\citeauthoryear{{Pawlowski}, {Famaey}, {Merritt}  \&
  {Kroupa}}{{Pawlowski} et~al.}{2015}]{Pawlowski15}
{Pawlowski} M.~S.,  {Famaey} B.,  {Merritt} D.,   {Kroupa} P.,  2015, \mn@doi
  [\apj] {10.1088/0004-637X/815/1/19}, \href
  {http://adsabs.harvard.edu/abs/2015ApJ...815...19P} {815, 19}

\bibitem[\protect\citeauthoryear{{Planck Collaboration} et~al.,}{{Planck
  Collaboration} et~al.}{2013}]{Planck2013}
{Planck Collaboration} et~al., 2013, arXiv:1303.5076, \href
  {http://adsabs.harvard.edu/abs/2013arXiv1303.5076P} {}

\bibitem[\protect\citeauthoryear{{Planck Collaboration} et~al.,}{{Planck
  Collaboration} et~al.}{2015}]{Planck2015}
{Planck Collaboration} et~al., 2015, arXiv:1502.01589, \href
  {http://adsabs.harvard.edu/abs/2015arXiv150201589P} {}

\bibitem[\protect\citeauthoryear{{Planck Collaboration} et~al.,}{{Planck
  Collaboration} et~al.}{2016}]{Planck2016}
{Planck Collaboration} et~al., 2016, preprint, \href
  {http://adsabs.harvard.edu/abs/2016arXiv160503507P} {} (\mn@eprint {arXiv}
  {1605.03507})

\bibitem[\protect\citeauthoryear{{Press} \& {Schechter}}{{Press} \&
  {Schechter}}{1974}]{PressSchechter}
{Press} W.~H.,  {Schechter} P.,  1974, \mn@doi [\apj] {10.1086/152650}, \href
  {http://adsabs.harvard.edu/abs/1974ApJ...187..425P} {187, 425}

\bibitem[\protect\citeauthoryear{{Rakavy} \& {Shaviv}}{{Rakavy} \&
  {Shaviv}}{1967}]{rs67}
{Rakavy} G.,  {Shaviv} G.,  1967, \mn@doi [\apj] {10.1086/149204}, \href
  {http://adsabs.harvard.edu/abs/1967ApJ...148..803R} {148, 803}

\bibitem[\protect\citeauthoryear{{Ritter}, {Safranek-Shrader}, {Gnat},
  {Milosavljevi{\'c}}  \& {Bromm}}{{Ritter} et~al.}{2012}]{ritt12}
{Ritter} J.~S.,  {Safranek-Shrader} C.,  {Gnat} O.,  {Milosavljevi{\'c}} M.,
  {Bromm} V.,  2012, \mn@doi [\apj] {10.1088/0004-637X/761/1/56}, \href
  {http://adsabs.harvard.edu/abs/2012ApJ...761...56R} {761, 56}

\bibitem[\protect\citeauthoryear{{Ritter}, {Safranek-Shrader}, {Milosavljevic}
  \& {Bromm}}{{Ritter} et~al.}{2016}]{Ritter16}
{Ritter} J.~S.,  {Safranek-Shrader} C.,  {Milosavljevic} M.,   {Bromm} V.,
  2016, preprint, \href {http://adsabs.harvard.edu/abs/2016arXiv160507236R} {}
  (\mn@eprint {arXiv} {1605.07236})

\bibitem[\protect\citeauthoryear{{Safranek-Shrader}, {Milosavljevi{\'c}}  \&
  {Bromm}}{{Safranek-Shrader} et~al.}{2014}]{ss13}
{Safranek-Shrader} C.,  {Milosavljevi{\'c}} M.,   {Bromm} V.,  2014, \mn@doi
  [\mnras] {10.1093/mnras/stt2307}, \href
  {http://adsabs.harvard.edu/abs/2014MNRAS.438.1669S} {438, 1669}

\bibitem[\protect\citeauthoryear{{Sasaki}, {Clark}, {Springel}, {Klessen}  \&
  {Glover}}{{Sasaki} et~al.}{2014}]{Sasaki2014}
{Sasaki} M.,  {Clark} P.~C.,  {Springel} V.,  {Klessen} R.~S.,   {Glover}
  S.~C.~O.,  2014, \mn@doi [\mnras] {10.1093/mnras/stu985}, \href
  {http://adsabs.harvard.edu/abs/2014MNRAS.442.1942S} {442, 1942}

\bibitem[\protect\citeauthoryear{{Schaerer}}{{Schaerer}}{2002}]{Schaerer2002}
{Schaerer} D.,  2002, \mn@doi [\aap] {10.1051/0004-6361:20011619}, \href
  {http://adsabs.harvard.edu/abs/2002A%26A...382...28S} {382, 28}

\bibitem[\protect\citeauthoryear{{Schauer}, {Whalen}, {Glover}  \&
  {Klessen}}{{Schauer} et~al.}{2015}]{Schauer2015}
{Schauer} A.~T.~P.,  {Whalen} D.~J.,  {Glover} S.~C.~O.,   {Klessen} R.~S.,
  2015, \mn@doi [\mnras] {10.1093/mnras/stv2117}, \href
  {http://ads.ari.uni-heidelberg.de/abs/2015MNRAS.454.2441S} {454, 2441}

\bibitem[\protect\citeauthoryear{{Schleicher}, {Galli}, {Glover}, {Banerjee},
  {Palla}, {Schneider}  \& {Klessen}}{{Schleicher} et~al.}{2009}]{Schleicher09}
{Schleicher} D.~R.~G.,  {Galli} D.,  {Glover} S.~C.~O.,  {Banerjee} R.,
  {Palla} F.,  {Schneider} R.,   {Klessen} R.~S.,  2009, \mn@doi [\apj]
  {10.1088/0004-637X/703/1/1096}, \href
  {http://adsabs.harvard.edu/abs/2009ApJ...703.1096S} {703, 1096}

\bibitem[\protect\citeauthoryear{{Sheth}, {Mo}  \& {Tormen}}{{Sheth}
  et~al.}{2001}]{ShethMoTormen}
{Sheth} R.~K.,  {Mo} H.~J.,   {Tormen} G.,  2001, \mn@doi [\mnras]
  {10.1046/j.1365-8711.2001.04006.x}, \href
  {http://adsabs.harvard.edu/abs/2001MNRAS.323....1S} {323, 1}

\bibitem[\protect\citeauthoryear{{Smidt}, {Whalen}, {Wiggins}, {Even},
  {Johnson}  \& {Fryer}}{{Smidt} et~al.}{2014}]{smidt13a}
{Smidt} J.,  {Whalen} D.~J.,  {Wiggins} B.~K.,  {Even} W.,  {Johnson} J.~L.,
  {Fryer} C.~L.,  2014, \mn@doi [\apj] {10.1088/0004-637X/797/2/97}, \href
  {http://adsabs.harvard.edu/abs/2014ApJ...797...97S} {797, 97}

\bibitem[\protect\citeauthoryear{{Smidt}, {Whalen}, {Chatzopoulos}, {Wiggins},
  {Chen}, {Kozyreva}  \& {Even}}{{Smidt} et~al.}{2015}]{smidt14a}
{Smidt} J.,  {Whalen} D.~J.,  {Chatzopoulos} E.,  {Wiggins} B.,  {Chen} K.-J.,
  {Kozyreva} A.,   {Even} W.,  2015, \mn@doi [\apj]
  {10.1088/0004-637X/805/1/44}, \href
  {http://adsabs.harvard.edu/abs/2015ApJ...805...44S} {805, 44}

\bibitem[\protect\citeauthoryear{{Smith} \& {Sigurdsson}}{{Smith} \&
  {Sigurdsson}}{2007}]{ss07}
{Smith} B.~D.,  {Sigurdsson} S.,  2007, \mn@doi [\apjl] {10.1086/518692}, \href
  {http://adsabs.harvard.edu/abs/2007ApJ...661L...5S} {661, L5}

\bibitem[\protect\citeauthoryear{{Smith} et~al.,}{{Smith}
  et~al.}{2007}]{nsmith07b}
{Smith} N.,  et~al., 2007, \mn@doi [\apj] {10.1086/519949}, \href
  {http://adsabs.harvard.edu/abs/2007ApJ...666.1116S} {666, 1116}

\bibitem[\protect\citeauthoryear{{Smith}, {Turk}, {Sigurdsson}, {O'Shea}  \&
  {Norman}}{{Smith} et~al.}{2009}]{bsmith09}
{Smith} B.~D.,  {Turk} M.~J.,  {Sigurdsson} S.,  {O'Shea} B.~W.,   {Norman}
  M.~L.,  2009, \mn@doi [\apj] {10.1088/0004-637X/691/1/441}, \href
  {http://adsabs.harvard.edu/abs/2009ApJ...691..441S} {691, 441}

\bibitem[\protect\citeauthoryear{{Smith}, {Glover}, {Clark}, {Greif}  \&
  {Klessen}}{{Smith} et~al.}{2011}]{sm11}
{Smith} R.~J.,  {Glover} S.~C.~O.,  {Clark} P.~C.,  {Greif} T.,   {Klessen}
  R.~S.,  2011, \mn@doi [\mnras] {10.1111/j.1365-2966.2011.18659.x}, \href
  {http://adsabs.harvard.edu/abs/2011MNRAS.414.3633S} {414, 3633}

\bibitem[\protect\citeauthoryear{{Smith}, {Wise}, {O'Shea}, {Norman}  \&
  {Khochfar}}{{Smith} et~al.}{2015}]{bsmith15}
{Smith} B.~D.,  {Wise} J.~H.,  {O'Shea} B.~W.,  {Norman} M.~L.,   {Khochfar}
  S.,  2015, \mn@doi [\mnras] {10.1093/mnras/stv1509}, \href
  {http://adsabs.harvard.edu/abs/2015MNRAS.452.2822S} {452, 2822}

\bibitem[\protect\citeauthoryear{{Spergel} et~al.,}{{Spergel}
  et~al.}{2015}]{wfirst}
{Spergel} D.,  et~al., 2015, arXiv:1503.03757

\bibitem[\protect\citeauthoryear{{Springel} et~al.,}{{Springel}
  et~al.}{2005}]{Millennium}
{Springel} V.,  et~al., 2005, \mn@doi [\nat] {10.1038/nature03597}, \href
  {http://adsabs.harvard.edu/abs/2005Natur.435..629S} {435, 629}

\bibitem[\protect\citeauthoryear{{Stacy} \& {Bromm}}{{Stacy} \&
  {Bromm}}{2013}]{sb13}
{Stacy} A.,  {Bromm} V.,  2013, \mn@doi [\mnras] {10.1093/mnras/stt789}, \href
  {http://adsabs.harvard.edu/abs/2013MNRAS.433.1094S} {433, 1094}

\bibitem[\protect\citeauthoryear{{Stacy}, {Greif}  \& {Bromm}}{{Stacy}
  et~al.}{2010}]{stacy10}
{Stacy} A.,  {Greif} T.~H.,   {Bromm} V.,  2010, \mn@doi [\mnras]
  {10.1111/j.1365-2966.2009.16113.x}, \href
  {http://adsabs.harvard.edu/abs/2010MNRAS.403...45S} {403, 45}

\bibitem[\protect\citeauthoryear{{Stacy}, {Bromm}  \& {Loeb}}{{Stacy}
  et~al.}{2011}]{Stacy2011}
{Stacy} A.,  {Bromm} V.,   {Loeb} A.,  2011, \mn@doi [\apjl]
  {10.1088/2041-8205/730/1/L1}, \href
  {http://adsabs.harvard.edu/abs/2011ApJ...730L...1S} {730, L1}

\bibitem[\protect\citeauthoryear{{Stacy}, {Greif}  \& {Bromm}}{{Stacy}
  et~al.}{2012}]{stacy12}
{Stacy} A.,  {Greif} T.~H.,   {Bromm} V.,  2012, \mn@doi [\mnras]
  {10.1111/j.1365-2966.2012.20605.x}, \href
  {http://adsabs.harvard.edu/abs/2012MNRAS.422..290S} {422, 290}

\bibitem[\protect\citeauthoryear{{Stacy}, {Bromm}  \& {Lee}}{{Stacy}
  et~al.}{2016}]{Stacy16}
{Stacy} A.,  {Bromm} V.,   {Lee} A.~T.,  2016, preprint, \href
  {http://adsabs.harvard.edu/abs/2016arXiv160309475S} {} (\mn@eprint {arXiv}
  {1603.09475})

\bibitem[\protect\citeauthoryear{{Susa}}{{Susa}}{2013}]{susa13}
{Susa} H.,  2013, \mn@doi [\apj] {10.1088/0004-637X/773/2/185}, \href
  {http://adsabs.harvard.edu/abs/2013ApJ...773..185S} {773, 185}

\bibitem[\protect\citeauthoryear{{Tamai} \& {Spyromilio}}{{Tamai} \&
  {Spyromilio}}{2014}]{EELT14}
{Tamai} R.,  {Spyromilio} J.,  2014, in Ground-based and Airborne Telescopes V.
  p. 91451E, \mn@doi{10.1117/12.2058467}

\bibitem[\protect\citeauthoryear{{Tanaka} \& {Haiman}}{{Tanaka} \&
  {Haiman}}{2009}]{th09}
{Tanaka} T.,  {Haiman} Z.,  2009, \mn@doi [\apj]
  {10.1088/0004-637X/696/2/1798}, \href
  {http://adsabs.harvard.edu/abs/2009ApJ...696.1798T} {696, 1798}

\bibitem[\protect\citeauthoryear{{Tegmark}, {Silk}, {Rees}, {Blanchard}, {Abel}
   \& {Palla}}{{Tegmark} et~al.}{1997}]{Tegmark97}
{Tegmark} M.,  {Silk} J.,  {Rees} M.~J.,  {Blanchard} A.,  {Abel} T.,   {Palla}
  F.,  1997, \mn@doi [\apj] {10.1086/303434}, \href
  {http://adsabs.harvard.edu/abs/1997ApJ...474....1T} {474, 1}

\bibitem[\protect\citeauthoryear{{Turk}, {Abel}  \& {O'Shea}}{{Turk}
  et~al.}{2009}]{turk09}
{Turk} M.~J.,  {Abel} T.,   {O'Shea} B.,  2009, \mn@doi [Science]
  {10.1126/science.1173540}, \href
  {http://adsabs.harvard.edu/abs/2009Sci...325..601T} {325, 601}

\bibitem[\protect\citeauthoryear{{Viel}, {Becker}, {Bolton}  \&
  {Haehnelt}}{{Viel} et~al.}{2013}]{Viel2013}
{Viel} M.,  {Becker} G.~D.,  {Bolton} J.~S.,   {Haehnelt} M.~G.,  2013, \mn@doi
  [\prd] {10.1103/PhysRevD.88.043502}, \href
  {http://adsabs.harvard.edu/abs/2013PhRvD..88d3502V} {88, 043502}

\bibitem[\protect\citeauthoryear{{Visbal}, {Haiman}  \& {Bryan}}{{Visbal}
  et~al.}{2015}]{Visbal2015}
{Visbal} E.,  {Haiman} Z.,   {Bryan} G.~L.,  2015, \mn@doi [\mnras]
  {10.1093/mnras/stv1941}, \href
  {http://ads.ari.uni-heidelberg.de/abs/2015MNRAS.453.4456V} {453, 4456}

\bibitem[\protect\citeauthoryear{{Volonteri}}{{Volonteri}}{2012}]{vol12}
{Volonteri} M.,  2012, \mn@doi [Science] {10.1126/science.1220843}, \href
  {http://adsabs.harvard.edu/abs/2012Sci...337..544V} {337, 544}

\bibitem[\protect\citeauthoryear{{Volonteri} \& {Gnedin}}{{Volonteri} \&
  {Gnedin}}{2009}]{Volonteri2009}
{Volonteri} M.,  {Gnedin} N.~Y.,  2009, \mn@doi [\apj]
  {10.1088/0004-637X/703/2/2113}, \href
  {http://adsabs.harvard.edu/abs/2009ApJ...703.2113V} {703, 2113}

\bibitem[\protect\citeauthoryear{{Whalen}}{{Whalen}}{2013}]{dw12}
{Whalen} D.~J.,  2013, Acta Polytechnica, \href
  {http://adsabs.harvard.edu/abs/2013AcPol..53..573W} {53, 573}

\bibitem[\protect\citeauthoryear{{Whalen} \& {Fryer}}{{Whalen} \&
  {Fryer}}{2012}]{wf12}
{Whalen} D.~J.,  {Fryer} C.~L.,  2012, \mn@doi [\apjl]
  {10.1088/2041-8205/756/1/L19}, \href
  {http://adsabs.harvard.edu/abs/2012ApJ...756L..19W} {756, L19}

\bibitem[\protect\citeauthoryear{{Whalen}, {Abel}  \& {Norman}}{{Whalen}
  et~al.}{2004}]{wan04}
{Whalen} D.,  {Abel} T.,   {Norman} M.~L.,  2004, \mn@doi [\apj]
  {10.1086/421548}, \href {http://adsabs.harvard.edu/abs/2004ApJ...610...14W}
  {610, 14}

\bibitem[\protect\citeauthoryear{{Whalen}, {O'Shea}, {Smidt}  \&
  {Norman}}{{Whalen} et~al.}{2008a}]{wet08b}
{Whalen} D.,  {O'Shea} B.~W.,  {Smidt} J.,   {Norman} M.~L.,  2008a, \mn@doi
  [\apj] {10.1086/587731}, \href
  {http://adsabs.harvard.edu/abs/2008ApJ...679..925W} {679, 925}

\bibitem[\protect\citeauthoryear{{Whalen}, {van Veelen}, {O'Shea}  \&
  {Norman}}{{Whalen} et~al.}{2008b}]{wet08a}
{Whalen} D.,  {van Veelen} B.,  {O'Shea} B.~W.,   {Norman} M.~L.,  2008b,
  \mn@doi [\apj] {10.1086/589643}, \href
  {http://adsabs.harvard.edu/abs/2008ApJ...682...49W} {682, 49}

\bibitem[\protect\citeauthoryear{{Whalen}, {Hueckstaedt}  \&
  {McConkie}}{{Whalen} et~al.}{2010}]{wet10}
{Whalen} D.,  {Hueckstaedt} R.~M.,   {McConkie} T.~O.,  2010, \mn@doi [\apj]
  {10.1088/0004-637X/712/1/101}, \href
  {http://adsabs.harvard.edu/abs/2010ApJ...712..101W} {712, 101}

\bibitem[\protect\citeauthoryear{{Whalen}, {Fryer}, {Holz}, {Heger}, {Woosley},
  {Stiavelli}, {Even}  \& {Frey}}{{Whalen} et~al.}{2013a}]{wet12a}
{Whalen} D.~J.,  {Fryer} C.~L.,  {Holz} D.~E.,  {Heger} A.,  {Woosley} S.~E.,
  {Stiavelli} M.,  {Even} W.,   {Frey} L.~H.,  2013a, \mn@doi [\apjl]
  {10.1088/2041-8205/762/1/L6}, \href
  {http://adsabs.harvard.edu/abs/2013ApJ...762L...6W} {762, L6}

\bibitem[\protect\citeauthoryear{{Whalen}, {Joggerst}, {Fryer}, {Stiavelli},
  {Heger}  \& {Holz}}{{Whalen} et~al.}{2013b}]{wet12c}
{Whalen} D.~J.,  {Joggerst} C.~C.,  {Fryer} C.~L.,  {Stiavelli} M.,  {Heger}
  A.,   {Holz} D.~E.,  2013b, \mn@doi [\apj] {10.1088/0004-637X/768/1/95},
  \href {http://adsabs.harvard.edu/abs/2013ApJ...768...95W} {768, 95}

\bibitem[\protect\citeauthoryear{{Whalen} et~al.,}{{Whalen}
  et~al.}{2013c}]{Whalen13c}
{Whalen} D.~J.,  et~al., 2013c, \mn@doi [\apj] {10.1088/0004-637X/768/2/195},
  \href {http://ads.ari.uni-heidelberg.de/abs/2013ApJ...768..195W} {768, 195}

\bibitem[\protect\citeauthoryear{{Whalen} et~al.,}{{Whalen}
  et~al.}{2013d}]{wet12b}
{Whalen} D.~J.,  et~al., 2013d, \mn@doi [\apj] {10.1088/0004-637X/777/2/110},
  \href {http://adsabs.harvard.edu/abs/2013ApJ...777..110W} {777, 110}

\bibitem[\protect\citeauthoryear{{Whalen}, {Smidt}, {Even}, {Woosley}, {Heger},
  {Stiavelli}  \& {Fryer}}{{Whalen} et~al.}{2014}]{wet13d}
{Whalen} D.~J.,  {Smidt} J.,  {Even} W.,  {Woosley} S.~E.,  {Heger} A.,
  {Stiavelli} M.,   {Fryer} C.~L.,  2014, \mn@doi [\apj]
  {10.1088/0004-637X/781/2/106}, \href
  {http://adsabs.harvard.edu/abs/2014ApJ...781..106W} {781, 106}

\bibitem[\protect\citeauthoryear{{Wise} \& {Abel}}{{Wise} \&
  {Abel}}{2005}]{wa05}
{Wise} J.~H.,  {Abel} T.,  2005, \mn@doi [\apj] {10.1086/430434}, \href
  {http://adsabs.harvard.edu/abs/2005ApJ...629..615W} {629, 615}

\bibitem[\protect\citeauthoryear{{Wise}, {Abel}, {Turk}, {Norman}  \&
  {Smith}}{{Wise} et~al.}{2012a}]{wise12a}
{Wise} J.~H.,  {Abel} T.,  {Turk} M.~J.,  {Norman} M.~L.,   {Smith} B.~D.,
  2012a, \mn@doi [\mnras] {10.1111/j.1365-2966.2012.21809.x}, \href
  {http://adsabs.harvard.edu/abs/2012MNRAS.427..311W} {427, 311}

\bibitem[\protect\citeauthoryear{{Wise}, {Turk}, {Norman}  \& {Abel}}{{Wise}
  et~al.}{2012b}]{wise12}
{Wise} J.~H.,  {Turk} M.~J.,  {Norman} M.~L.,   {Abel} T.,  2012b, \mn@doi
  [\apj] {10.1088/0004-637X/745/1/50}, \href
  {http://adsabs.harvard.edu/abs/2012ApJ...745...50W} {745, 50}

\bibitem[\protect\citeauthoryear{{Woosley}, {Blinnikov}  \& {Heger}}{{Woosley}
  et~al.}{2007}]{wbh07}
{Woosley} S.~E.,  {Blinnikov} S.,   {Heger} A.,  2007, \mn@doi [\nat]
  {10.1038/nature06333}, \href
  {http://adsabs.harvard.edu/abs/2007Natur.450..390W} {450, 390}

\bibitem[\protect\citeauthoryear{{Xu}, {Wise}  \& {Norman}}{{Xu}
  et~al.}{2013}]{xu13}
{Xu} H.,  {Wise} J.~H.,   {Norman} M.~L.,  2013, \mn@doi [\apj]
  {10.1088/0004-637X/773/2/83}, \href
  {http://adsabs.harvard.edu/abs/2013ApJ...773...83X} {773, 83}

\bibitem[\protect\citeauthoryear{{Xu}, {Ahn}, {Wise}, {Norman}  \&
  {O'Shea}}{{Xu} et~al.}{2014}]{Xu2014}
{Xu} H.,  {Ahn} K.,  {Wise} J.~H.,  {Norman} M.~L.,   {O'Shea} B.~W.,  2014,
  \mn@doi [\apj] {10.1088/0004-637X/791/2/110}, \href
  {http://adsabs.harvard.edu/abs/2014ApJ...791..110X} {791, 110}

\bibitem[\protect\citeauthoryear{{Xu}, {Wise}, {Norman}, {Ahn}  \&
  {O'Shea}}{{Xu} et~al.}{2016}]{XuNorm}
{Xu} H.,  {Wise} J.~H.,  {Norman} M.~L.,  {Ahn} K.,   {O'Shea} B.~W.,  2016,
  preprint, \href {http://adsabs.harvard.edu/abs/2016arXiv160407842X} {}
  (\mn@eprint {arXiv} {1604.07842})

\bibitem[\protect\citeauthoryear{{Yoon}, {Dierks}  \& {Langer}}{{Yoon}
  et~al.}{2012}]{yoon12}
{Yoon} S.-C.,  {Dierks} A.,   {Langer} N.,  2012, \mn@doi [\aap]
  {10.1051/0004-6361/201117769}, \href
  {http://adsabs.harvard.edu/abs/2012A%26A...542A.113Y} {542, A113}

\bibitem[\protect\citeauthoryear{{Yoshida}, {Abel}, {Hernquist}  \&
  {Sugiyama}}{{Yoshida} et~al.}{2003}]{Yoshida2003}
{Yoshida} N.,  {Abel} T.,  {Hernquist} L.,   {Sugiyama} N.,  2003, \mn@doi
  [\apj] {10.1086/375810}, \href
  {http://adsabs.harvard.edu/abs/2003ApJ...592..645Y} {592, 645}

\bibitem[\protect\citeauthoryear{{Yoshida}, {Omukai}, {Hernquist}  \&
  {Abel}}{{Yoshida} et~al.}{2006}]{yoha06}
{Yoshida} N.,  {Omukai} K.,  {Hernquist} L.,   {Abel} T.,  2006, \mn@doi [\apj]
  {10.1086/507978}, \href {http://adsabs.harvard.edu/abs/2006ApJ...652....6Y}
  {652, 6}

\bibitem[\protect\citeauthoryear{{Yoshida}, {Omukai}  \& {Hernquist}}{{Yoshida}
  et~al.}{2008}]{y08}
{Yoshida} N.,  {Omukai} K.,   {Hernquist} L.,  2008, \mn@doi [Science]
  {10.1126/science.1160259}, \href
  {http://adsabs.harvard.edu/abs/2008Sci...321..669Y} {321, 669}

\bibitem[\protect\citeauthoryear{{de Souza}, {Ishida}, {Johnson}, {Whalen}  \&
  {Mesinger}}{{de Souza} et~al.}{2013}]{ds13}
{de Souza} R.~S.,  {Ishida} E.~E.~O.,  {Johnson} J.~L.,  {Whalen} D.~J.,
  {Mesinger} A.,  2013, \mn@doi [\mnras] {10.1093/mnras/stt1680}, \href
  {http://adsabs.harvard.edu/abs/2013MNRAS.436.1555D} {436, 1555}

\bibitem[\protect\citeauthoryear{{de Souza}, {Ishida}, {Whalen}, {Johnson}  \&
  {Ferrara}}{{de Souza} et~al.}{2014}]{ds14}
{de Souza} R.~S.,  {Ishida} E.~E.~O.,  {Whalen} D.~J.,  {Johnson} J.~L.,
  {Ferrara} A.,  2014, \mn@doi [\mnras] {10.1093/mnras/stu984}, \href
  {http://adsabs.harvard.edu/abs/2014MNRAS.442.1640D} {442, 1640}

\makeatother
\end{thebibliography}
\end{document}